\documentclass[11pt]{article}
\usepackage{jcappub}
\usepackage{bm}
\usepackage{color}
\usepackage{graphicx}
\usepackage{multirow}
\usepackage{enumitem}
\usepackage{cprotect}
\usepackage{amssymb}
\usepackage[normalem]{ulem}
\usepackage{fancyvrb}
\usepackage{upquote}
\usepackage{multirow}
\definecolor{ourcolor}{rgb}{0.7, 0.25, 0.05}

\newcommand{\qmsq}{\qmsq}
\renewcommand{\qmsq}{\q^2}

\newcommand{\be}{\begin{equation}}
\newcommand{\ee}{\end{equation}}
\newcommand{\een}{\end{subequations}}
\newcommand{\ben}{\begin{subequations}}
\newcommand{\beq}{\begin{eqalignno}}
\newcommand{\eeq}{\end{eqalignno}}

\newcommand{\lsim}{\mathrel{\mathop{\kern 0pt \rlap
      {\raise.2ex\hbox{$<$}}}\lower.9ex\hbox{\kern-.190em $ \sim$}}}
\newcommand{\gsim}{\mathrel{\mathop{\kern 0pt
      \rlap{\raise.2ex\hbox{$>$}}}\lower.9ex\hbox{\kern-.190em $\sim$}}}

\newcommand{\VectorTypefaceArrow}{
\let\oldvec\vec
\renewcommand{\vec}[1]{\oldvec{##1}} 
\newcommand{\uvec}[1]{\hat{##1}} 
}

\VectorTypefaceArrow

\usepackage{upgreek} 
\newcommand{\q}{{\widetilde{q}}}
\newcommand{\calO}{{\cal O}}

\usepackage{mathtools}

\newcommand{\plus}{\,+}

\usepackage{booktabs}
\usepackage{xcolor}
\usepackage{arydshln}

\allowdisplaybreaks

\title{\includegraphics[width=5cm]{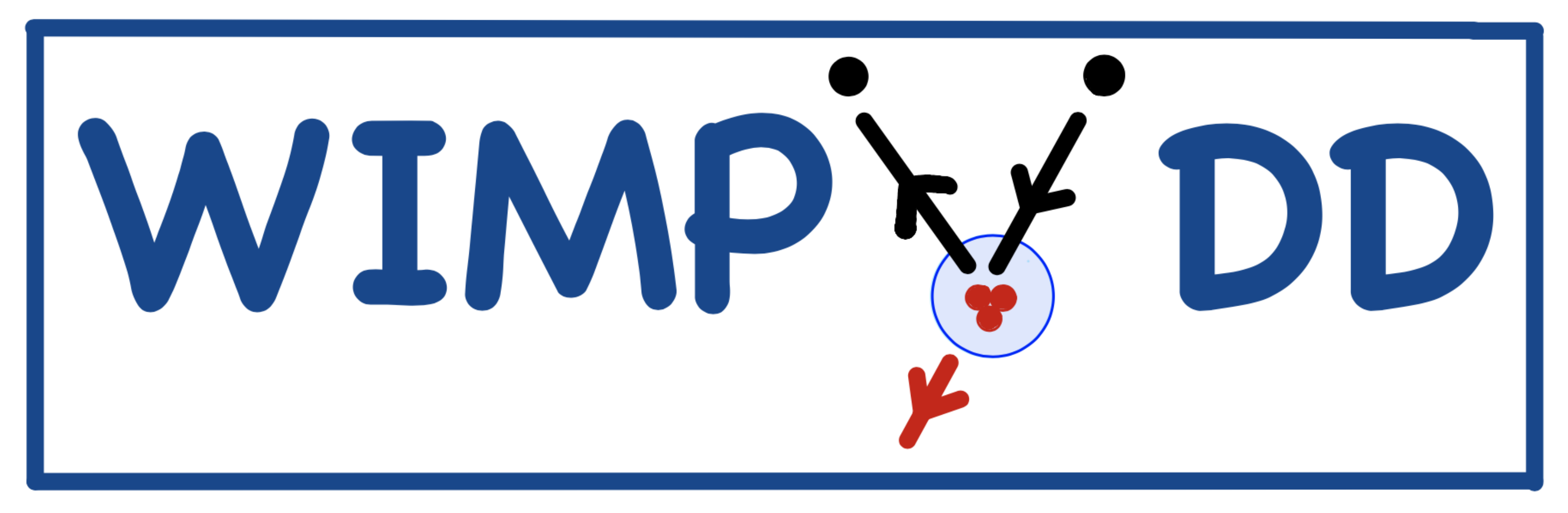}\\[0.3cm]WimPyDD: an object--oriented Python code for the calculation of WIMP direct detection signals}

\author[a]{Injun Jeong,}
\author[a]{Sunghyun Kang,}
\author[a]{Stefano Scopel,}
\author[b]{Gaurav Tomar}
\affiliation[a]{Department of Physics, Sogang University, Seoul 121-742, South Korea}
\affiliation[b]{Physik-Department, Technische Universit\"at M\"unchen, James-Franck-Stra\ss e, 85748
Garching, Germany}
\emailAdd{natson@naver.com}
\emailAdd{francis735@naver.com}
\emailAdd{scopel@sogang.ac.kr}
\emailAdd{physics.tomar@tum.de}

\abstract{We introduce WimPyDD, a modular, object--oriented and customizable
Python code that calculates accurate predictions for the expected
rates in Weakly Interacting Massive Particle (WIMP) direct--detection experiments within the framework of
Galilean--invariant non--relativistic effective theory in virtually
any scenario, including inelastic scattering, an arbitrary WIMP spin
and a generic WIMP velocity distribution in the Galactic halo.
WimPyDD exploits the factorization of the three main components that
enter in the calculation of direct detection signals: i) the Wilson
coefficients that encode the dependence of
the signals on the ultraviolet completion of the effective theory; ii)
a response function that depends on the nuclear physics and 
on the main features of the experimental detector (acceptance, energy resolution,
response to nuclear recoils); iii) a halo function that depends on the
WIMP velocity distribution and that encodes the astrophysical inputs.
In WimPyDD these three components are calculated and stored separately
for later interpolation and combined together only as the last step
of the signal evaluation procedure. This makes the phenomenological
study of the direct detection scattering rate with WimPyDD transparent
and fast also when the parameter space of the WIMP model has a large
dimensionality.}

\begin{document}
\hspace*{107.5mm}{CQUeST-2021-0663}\\
\hspace*{116mm}{TUM-HEP 1343/21}
\maketitle
\section{Introduction}
\label{sec:introduction}
Weakly Interacting Massive Particles (WIMPs) provide the most popular
candidates for the Cold Dark Matter (CDM) that is known from
gravitational measurements to constitute about 25\% of the density of
the Universe, is concentrated in Galaxies like the Milky Way and is
required in any model of Galaxy formation. WIMPs have feeble
interactions of the Weak or Super-Weak type with ordinary matter that
allow them to thermalize in the early Universe and that can predict
the correct relic density through the thermal decoupling mechanism. A
worldwide experimental effort is under way to observe such
interactions. In direct detection (DD) experiments WIMP-nucleus
scattering events are searched for in solid-state or liquid-state
detectors in underground laboratories protected by cosmic rays.

The calculation of DD signals depends on particle
physics, nuclear physics and astrophysics inputs. Moreover, in order
to obtain realistic predictions, experimental features such as the
energy resolution and the efficiency of the detector and the specific
response of the target material to nuclear recoils (quenching, light
yield) are required. In the present paper we introduce WimPyDD, a
customizable, object--oriented Python code that allows to handle all
such different inputs in a modular way to obtain accurate predictions
for the expected signal in WIMP DD experiments in
virtually any theoretical scenario. Before its present release WimPyDD
was extensively tested in several
publications~\cite{sogang_scaling_law_nr, sogang_scaling_law_rel,
  cosine100_dama,dama_2018_sogang,dama_psidm_viable,dama_inelastic_eft_sogang,
  all_spin_pheno}.

In particular, WimPyDD is specifically designed to calculate WIMP DD signals in 
generalized scenarios, whose relevance in present research is twofold: first, 
accelerator physics has not provided so far any
information or hint about the physics beyond the standard model that
must be related to the Dark Matter (DM) particle, so that
model-independent approaches have become increasingly popular to
interpret Dark Matter search experiments.
Moreover, the incoming WIMP flux in DD experiments
depends on the velocity distribution of such particles in the halo of
our Galaxy, which is in principle not known.

Since the DD process is non--relativistic (NR) the WIMP-nucleon
interaction can be parameterized with an effective Hamiltonian
${\bf\mathcal{H}}$ that complies with Galilean symmetry.  The
effective Hamiltonian ${\bf\mathcal{H}}$ to zero-th order in the
WIMP-nucleon relative velocity $\vec{v}$ and momentum transfer
$\vec{q}$ has been known since at least Ref.~\cite{GoodmanWitten}, and
consists of the usual spin-dependent (SD) and spin-independent (SI)
terms.  To first order in $\vec{v}$, the effective Hamiltonian
${\bf\mathcal{H}}$ has been systematically described
in~\cite{haxton1,haxton2} for WIMPs of spin 0 and 1/2, and less
systematically described in~\cite{krauss_spin_1,catena_krauss_spin_1}
for WIMPs of spin 1 and in~\cite{barger_2008} for WIMPs of spin 3/2.
Recently in~\cite{all_spins} the NR
effective Hamiltonian for WIMP--nucleous scattering has been extended
to include WIMPs of arbitrary spin $j_\chi$.  
In Ref.~\cite{all_spins} ${\bf\mathcal{H}}$ is
written in a complete base of rotationally invariant operators
organized according to the rank of the $2 j_\chi+1$ irreducible
operator products of up to $2 j_\chi$ WIMP spin vectors.  In
particular, for a WIMP of spin $j_\chi$ a base of $4+20 j_\chi$
independent operators is obtained that can be matched to any high-energy
model of particle dark matter, including elementary particles and
composite states. WimPyDD incorporates the formalism of  Ref.~\cite{all_spins}, allowing
to introduce an effective Hamiltonian for generic WIMP spin $j_\chi$
and Wilson coefficients that are arbitrary functions of external
parameters and of the momentum $q$ transferred in the scattering
process.

To calculate expected signals WimPyDD makes full use of the fact that
DD signals can be written as the convolution of three factorized
components: i) the Wilson coefficients, that depend on arbitrary
external parameters $w_i$; ii) a response function ${\cal R}$, that
depends on the nuclear physics and on the features of the experimental
detector (acceptance, energy resolution, response to nuclear recoils);
iii) a halo function $\eta$, that depends on the WIMP velocity
distribution and that encodes the astrophysical
inputs~\cite{factorization, generalized_halo_indep}.  In WimPyDD these
three components are calculated separately and combined together only
at the last step of the signal evaluation procedure.  In particular
the response functions ${\cal R}$ and the halo function $\eta$ are
calculated and stored for later interpolation using parameterizations
in terms of the recoil energy and the WIMP incoming velocity,
respectively, that do not depend on the scattering kinematics (i.e. on
the WIMP mass $m_\chi$ and on the mass splitting $\delta$ in the case
of inelastic scattering (IDM)~\cite{inelastic}\footnote{
In IDM a DM
 particle  of mass $m_{\chi}$ interacts with
 atomic nuclei exclusively by either up--scattering to a heavier
 state with mass $m_{\chi}+\delta$ or down--scattering to a lighter 
 state with mass $m_{\chi}-\delta$ (this latter case is referred to as exothermic DM).  
 WimPyDD can handle both cases.} ). This
makes the phenomenological study of the DD scattering rate transparent
and fast also when the parameter space $(m_\chi,\delta,w_i)$ has a
large dimensionality.

The paper is organized as follows: in Section~\ref{sec:theory} we
introduce the theoretical background of WIMP nuclear scattering and
show the factorizations used by WimPyDD to optimize the expressions
used for the numerical evaluation of the signals; in
Section~\ref{sec:thecode} we introduce the code. We explain its
installation in~\ref{sec:quick_start}; introduce the signal routines
in~\ref{sec:signal_routines} and the classes required to use them
in~\ref{sec:classes}. At the core of how WimPyDD works are the
integrated response functions ${\cal R}$, which at the stage of signal evaluation
are {\it below the lid} and invisible to the user. However knowing how
such response functions are handled by WimPyDD is crucial for the user
to run the code properly and to exploit its optimizations, so
Section~\ref{sec:handling} provides a detailed discussion of this
issue, with a particular focus on the aspect of how Wilson
coefficients that depend on the exchanged momentum are implemented. In
Section~\ref{sec:examples} we illustrate the use of WimPyDD through
some examples. Several details needed to understand the code are
provided in the Appendices: Appendix~\ref{app:response_functions} integrates
Section~\ref{sec:theory} with detailed expressions for the scattering
rates and the response functions; Appendix~\ref{app:eft} illustrates the two
alternative bases for the operators of the effective Hamiltonian that
can be used in the code; in Appendix~\ref{app:exp} a detailed explanation of
how to initialize a DD experimental set--up in WimPyDD is given;
Appendix~\ref{app:add_elements} explains how to add new nuclear targets and
their response functions; Appendix ~\ref{sec:halo} explains how to calculate the
halo function; finally, Appendix ~\ref{app:summary} provides a summary of the 
WimPyDD components and their correspondence to the theoretical formulas
introduced in Section~\ref{sec:theory} and in Appendix~\ref{app:response_functions}.

\section{WIMP direct detection scattering rate}
\label{sec:theory}

Formulas for the calculation of expected WIMP DD signals
have been available in the literature for a long time (see for
instance~\cite{Lewin_Smith_1996}). However to optimize their numerical
evaluation WimPyDD makes explicit use of the factorization among
Wilson coefficients, the halo function and the experimental response
functions, and exploits a parameterization of the latter that is
independent on the WIMP mass $m_\chi$ and on the mass splitting
$\delta$. Since the relevant expressions are not available in the
literature in this Section we outline them (additional
details are provided in~\ref{app:response_functions}).

The expected number of events in a WIMP DD experiment in
the interval of visible energy $E_1^{\prime}\le E^{\prime}\le
E_2^{\prime}$ can be cast in the form~\cite{generalized_halo_indep}:

\begin{equation}
  R_{[E_1^{\prime},E_2^{\prime}]}(t)=\frac{\rho_\chi}{m_\chi}\int_{v_{T^*}}^{\infty}
  \sum_T dv {\cal R}_{T,[E_1^{\prime},E_2^{\prime}]}(v)\eta(v,t),
  \label{eq:r_eta}
  \end{equation}

\noindent with $\rho_\chi$ the WIMP local density in the neighborhood of the
Sun, $m_\chi$ the WIMP mass,

\begin{equation}
v_{T^*}\equiv\left\{\begin{aligned}
\sqrt{\frac{2\delta}{\mu_{\chi T}}} ,\quad
\text{if}\;\delta> 0\\
0,\quad
\text{if}\; \delta\le 0\, ,
\end{aligned}\right. 
\label{eq:vstar}
\end{equation}

\noindent and the sum is over the nuclear targets $T$.
In the expression above: 

\begin{equation}
 \eta(v,t)=\int_{v}^{\infty}\frac{f(\vec{v^{\prime}},t)}{|\vec{v}^{\prime}|}\,d^3v^{\prime},
 =\int_{v}^{\infty}\frac{f(v^{\prime},t)}{v^{\prime}}\,dv^{\prime},
\label{eq:eta}  
\end{equation}

\noindent is a halo function that depends on the WIMP speed
distribution $f(v,t)\equiv 1/(4\pi) \int d\Omega
{v^\prime}^2f(\vec{v}^\prime,t)$ (present DD experiments are not sensitive to
the incoming direction of the WIMP), while ${\cal
  R}^T_{[E_1^{\prime},E_2^{\prime}]}(v)$ is a response function that
depends on the WIMP--target interaction and on the features of the
detector.  If the speed distribution is written in terms of a
superposition of $N_s$ streams:

\begin{equation}
f(v,t)=\sum_k^{N_s} \lambda_k(t) \delta(v-v_k),
\end{equation}

\noindent the halo function becomes: 

\begin{equation}
  \eta(v,t)= \sum_{k=1}^{N_s}
  \frac{\lambda_k(t)}{v_k}\Theta(v_k-v)=\sum_{k=1}^{N_s}
  \delta\eta_k(t)\Theta(v_k-v),
\label{eq:delta_eta_piecewise}
\end{equation}  

\noindent and in Eq.~(\ref{eq:r_eta}) one has the following factorization:

\begin{equation}
  R_{[E_1^{\prime},E_2^{\prime}]}(t)=\frac{\rho_\chi}{m_\chi}\sum_{k=1}^{N_s}
  \delta\eta_k(t) \int_{v_{T^*}}^{v_i}
  \sum_T dv {\cal R}_{T,[E_1^{\prime},E_2^{\prime}]}(v).
  \label{eq:r_eta_streams}
  \end{equation}

\label{sec:streamed_halo_function}
Most DD experiments put upper bounds on the time average
of $R_{[E_1^{\prime},E_2^{\prime}]}(t)$:

\begin{equation}
S^{(0)}_{[E_1^{\prime},E_2^{\prime}]}\equiv \frac{1}{T_0}\int_0^{T_0}
R_{[E_1^{\prime},E_2^{\prime}]}(t)dt,
\label{eq:s0}
\end{equation}  

\noindent (with $T_0$=1 year) while dedicated
large--mass detectors such as those of DAMA~\cite{dama_2018}, 
COSINE~\cite{cosine_modulation_2019} or ANAIS~\cite{anais_modulation_2019} are
sensitive to the yearly modulation effect ensuing from the time
dependence of the Earth velocity $\vec{v}_E(t)$ due to its rotation around the
Sun. In particular, the latter measure the cosine transform of
$R_{[E_1^{\prime},E_2^{\prime}]}(t)$:

\begin{equation}
S^{(1)}_{[E_1^{\prime},E_2^{\prime}]}\equiv \frac{2}{T_0}\int_0^{T_0}
\cos\left[\frac{2\pi}{T_0}(t-t_0)\right]R_{[E_1^{\prime},E_2^{\prime}]}(t)dt,
\label{eq:sm}
\end{equation}  

\noindent with $t_0$=June, 2$^{nd}$. Using~Eq.(\ref{eq:r_eta})
in~Eq.(\ref{eq:s0}) and Eq.(\ref{eq:sm}) one gets:

\begin{eqnarray}
  S^{(0,1)}_{[E_1^{\prime},E_2^{\prime}]}&=&\int_{v^*_{T}}^{\infty} \sum_T dv {\cal R}_{T,[E_1^{\prime},E_2^{\prime}]}(v)\eta^{(0,1)}(v),\nonumber\\
   \eta^{(0)}(v)&=& \frac{1}{T_0}\int_0^{T_0} \eta(v,t), \label{eq:eta0}\\
  \eta^{(1)}(v)&=& \frac{2}{T_0}\int_0^{T_0}
\cos\left[\omega(t-t_0)\right] \eta(v,t),\,\,\,\,\omega=\frac{2\pi}{T_0}.
  \label{eq:eta1}
  \end{eqnarray}
\noindent where $\mu_{\chi T}$ is the WIMP--target reduced mass. This implies that Eq.~(\ref{eq:r_eta_streams}) can be
directly used to calculate $S^{(0)}_{[E_1^{\prime},E_2^{\prime}]}$ and
$S^{(1)}_{[E_1^{\prime},E_2^{\prime}]}$, provided that the substitution
$\delta\eta_k(t)\rightarrow \delta\eta_k^{(0,1)}$, with:

\begin{equation}
  \eta^{(0,1)}(v)= \sum_{k=1}^{N_s}
  \delta\eta_k^{(0,1)}\theta(v_k-v).
\label{eq:delta_eta01_piecewise}
\end{equation}

\noindent is made.

In Appendix~\ref{app:response_functions} (see Eq.(\ref{eq:proof_r_dv_de})) it
is shown that the integral on the right--hand side of
Eq.~(\ref{eq:r_eta_streams}) takes the form:

\begin{align}\nonumber
\int_{v_{T^*}}^{v_i} \sum_T dv {\cal R}_{T,[E_1^{\prime},E_2^{\prime}]}(v) &=& \\ \int_{E_R^{min}(v_i)}^{E_R^{max}(v_i)}\sum_T d E_R &\left[ {\cal R}^0_{T,[E_1^{\prime},E_2^{\prime}]}(E_R)
+ {\cal R}^1_{T,[E_1^{\prime},E_2^{\prime}]}(E_R)(v_i^2-v^2_{T,min}(E_R)) \right ],
\label{eq:r_dv_de}
  \end{align}

\noindent (see Eqs.~(\ref{eq:r_T_i}, \ref{eq:differential_r}) for the explicit expressions of ${\cal R}^0_{T,[E_1^{\prime},E_2^{\prime}]}$ and ${\cal R}^1_{T,[E_1^{\prime},E_2^{\prime}]}$). In Eq.~(\ref{eq:r_dv_de}):

\begin{equation}
  v_{T,min}(E_R)=\frac{1}{\sqrt{2 m_T E_R}}\left | \frac{m_TE_R}{\mu_{\chi T}}+\delta \right |,
  \label{eq:vmin}
\end{equation}

\noindent is the minimal speed an incoming WIMP needs to have in the
target reference frame to deposit energy $E_R$ (with $m_T$ the nuclear
target and $\delta$
the mass splitting in case of inelastic scattering).
In the same equation:

\begin{equation}
  E_R^{min,max}(v)=\left[\frac{\mu_{\chi T}}{\sqrt{2 m_T}}\left ( v\mp\sqrt{v^2-v_{T^*}^2}\right ) \right ]^2,
  \label{eq:er_max_min}
\end{equation}

\noindent with $ v_{T^*}$ given in Eq.~(\ref{eq:vstar}).

\noindent The most general WIMP-nucleus interaction complying with
Galilean symmetry can
be parameterized with an effective Hamiltonian ${\bf\mathcal{H}}$ of
the
form~\cite{haxton1,haxton2,krauss_spin_1,catena_krauss_spin_1,all_spins}:
\begin{eqnarray}
{\bf\mathcal{H}}(w_i,q)&=& \sum_{\tau=0,1} \sum_{j} c_j^{\tau}(w_i,q) \mathcal{O}_{j}({\bf{q}}) \, t^{\tau}.
\label{eq:H}
\end{eqnarray}

\noindent In the equation above we have explicitly written the
dependence of the effective Hamiltonian on some parameters $w_i$
(which can include the WIMP mass $m_\chi$ and the mass splitting
$\delta$) arising from the matching of the Wilson coefficients of the
low--energy effective theory to its ultraviolet completion, and on the
transferred momentum $q$ (for instance through a
propagator). All $\mathcal{O}_j$ operators for $j_\chi\le$ 1/2, and some for $j_\chi$ = 1 are listed in Table~\ref{tab:Haxton_operators}. Moreover, in Eq.~(\ref{eq:H}), $t^0=1$, $t^1=\tau_3$
denote the $2\times2$ identity and third Pauli matrix in the nucleon
isospin space, respectively, and the isoscalar and isovector Wilson
coefficients coupling $c^0_j$ and $c^{1}_j$ are related to those to
protons and neutrons $c^{p}_j$ and $c^{n}_j$ by
$c^{p}_j=(c^{0}_j+c^{1}_j)/2$ and $c^{n}_j=(c^{0}_j-c^{1}_j)/2$.

If the scattering process is driven by the Hamiltonian of
Eq.~(\ref{eq:H}) the functions ${\cal
  R}^i_{T,[E_1^{\prime},E_2^{\prime}]}$ in Eq.(\ref{eq:r_dv_de})
are quadratic in the Wilson coefficients of the effective theory, that
can be factored out:

\begin{eqnarray}
&&{\cal R}^i_{T,[E_1^{\prime},E_2^{\prime}]} (E_R)=\sum_{j,k}
c_j^{\tau}(w_i,q_0)c_{k}^{\tau^\prime}(w_i,q_0^{\prime})\left [{\cal R}^i_{T,[E_1^{\prime},E_2^{\prime}]}\right ]_{jk}^{\tau\tau^{\prime}}(E_R).
\label{eq:r_factorization}
  \end{eqnarray}

\noindent In the expression above $q_0$, $q_0^{\prime}$ are arbitrary
momentum scales in the case when the Wilson coefficients $c_j^{\tau}$,
$c_{k}^{\tau^\prime}$ have an explicit dependence on $q$. Using in Eq.~(\ref{eq:r_dv_de}) the explicit expression of the square of $v_{T,min}(E_R)$:

\begin{equation}
 v_{T,min}(E_R)^2=\frac{m_T}{2\mu_{\chi T}^2}E_R+\frac{\delta^2}{2 m_T}\frac{1}{E_R}+\frac{\delta}{\mu_{\chi T}}  
\label{eq:vmin2}
\end{equation}

\noindent and defining the integrated response functions:

\begin{eqnarray}
  &&\left [\bar{{\cal R}}^i_{T,[E_1^{\prime},E_2^{\prime}]}\right ]_{jk}^{\tau\tau^{\prime}}(E_R)\equiv \int_0^{E_R} dE^{\prime}_R \left[{\cal R}^i_{T,[E_1^{\prime},E_2^{\prime}]}\right ]_{jk}^{\tau\tau^{\prime}}(E_R^{\prime}),\,\,\,i=0,1 \nonumber\\
  &&\left [\bar{{\cal R}}^{1E}_{T,[E_1^{\prime},E_2^{\prime}]}\right ]_{jk}^{\tau\tau^{\prime}}(E_R)\equiv \int_0^{E_R} dE^{\prime}_R E_R^{\prime} \left [{\cal R}^{1}_{T, [E_1^{\prime},E_2^{\prime}]}\right ]_{jk}^{\tau\tau^{\prime}}(E_R^{\prime}) \nonumber\\
&&\left [\bar{{\cal R}}^{1E^{-1}}_{T, [E_1^{\prime},E_2^{\prime}]}\right]_{jk}^{\tau\tau^{\prime}}(E_R)\equiv \int_0^{E_R} dE^{\prime}_R \frac{1}{E_R^{\prime}} \left [{\cal R}^{1}_{T, [E_1^{\prime},E_2^{\prime}]}\right ]_{jk}^{\tau\tau^{\prime}}(E_R^{\prime}),\nonumber\\
  \label{eq:rbar}
\end{eqnarray}

\noindent the predicted rate for both the time-averaged and the yearly
modulated signals can then be written as:

\begin{eqnarray}
  &&S^{(0,1)}_{[E_1^{\prime},E_2^{\prime}]}=\frac{\rho_{\chi}}{m_{\chi}}\sum_{k=1}^{N_s} \delta\eta_k^{(0,1)}\times \sum_T\sum_{ij}\sum_{\tau\tau^{\prime}} c_j^{\tau}(w_i,q_0)c_{k}^{\tau^\prime}(w_i,q_0^{\prime})  \nonumber\\
  &&  \left \{ \left [\bar{{\cal R}}^0_{T,[E_1^{\prime},E_2^{\prime}]}\right ]_{jk}^{\tau\tau^{\prime}}(E_R)
  +(\frac{v_k^2}{c^2}-\frac{\delta}{\mu_{\chi T}})\left [
  \bar{{\cal R}}^1_{T,[E_1^{\prime},E_2^{\prime}]}\right ]_{jk}^{\tau\tau^{\prime}}(E_R) \right .\nonumber\\
&& \left .  -\frac{m_T}{2\mu_{\chi T}^2}
  \left [\bar{{\cal R}}^{1E}_{T,[E_1^{\prime},E_2^{\prime}]}\right ]_{jk}^{\tau\tau^{\prime}}(E_R)-\frac{\delta^2}{2 m_T}
  \left[\bar{{\cal R}}^{1E^{-1}}_{T,[E_1^{\prime},E_2^{\prime}]}\right ]_{jk}^{\tau\tau^{\prime}}(E_R)\right \}^{E_R^{max}(v_k)}_{E_R^{min}(v_k)},\nonumber \\
 && {} \label{eq:rate_rbar_inelastic}
\end{eqnarray}

\noindent where $\{f(E_R)\}_{E_1}^{E_2} \equiv f(E_2)-f(E_1)$.  The
expression above is the master formula used by WimPyDD to calculate
expected rates. As anticipated, it exploits explicitly the
factorization of the the Wilson coefficients $c_j^{\tau}$, the
response functions $\left [\bar{{\cal
      R}}^a_{T,[E_1^{\prime},E_2^{\prime}]}\right
]_{jk}^{\tau\tau^{\prime}}$ and of the the halo functions
$\delta\eta_k^{(0,1)}$. Crucially, in Eq.~(\ref{eq:rate_rbar_inelastic})
the only dependence on $m_\chi$ and $\delta$ is through
$E_R^{min,max}(v)$ (see Eq.~(\ref{eq:er_max_min})).
%
%
In particular, the numerical calculation of the integrated response functions $\left
[\bar{{\cal R}}^a_{T, [E_1^{\prime},E_2^{\prime}]}\right
]_{jk}^{\tau\tau^{\prime}}$ can be time consuming since they require
the evaluation of double integrals\footnote{Notice that a different
  response function must be calculated for each target, energy bin,
  $c_i^{\tau}c_j^{\tau^{\prime}}$ combination and
  $a=0,1,1E,1E^{-1}$.}. However, they depend only on the single
external argument $E_R$.  The main strategy used by WimPyDD to speed
up the calculation of the rates is to tabulate each $\left [\bar{{\cal
      R}}^a_{T,[E_1^{\prime},E_2^{\prime}]}\right
]_{jk}^{\tau\tau^{\prime}}$ as a function of $E_R$ for later
interpolation. As already mentioned such tables do not depend on the
WIMP mass $m_\chi$ and on the mass splitting $\delta$, that only enter
in the mapping between the $v_k$ values and the
energies $E_R^{max}(v_k)$ and $E_R^{min}(v_k)$ at which the response
functions in the tables are interpolated.

\section{The WimPyDD code}
\label{sec:thecode}
\begin{figure}
\begin{center}
  \includegraphics[width=0.9\columnwidth]{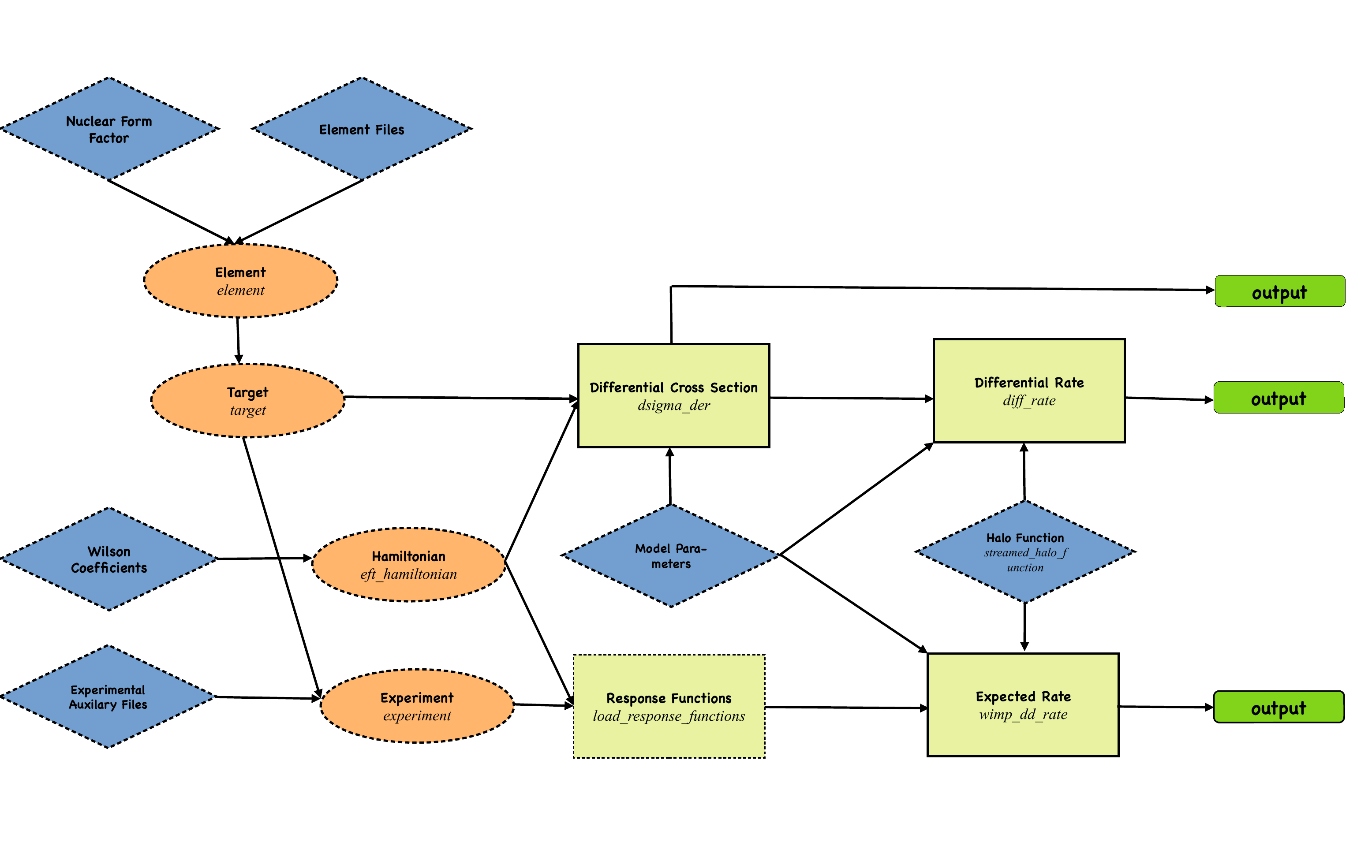}
  \end{center}
\caption{Main structure of WimPyDD. Diamond--shaped elements
indicate the input from the user, oval shapes contain classes and
rectangles represent functions. Three final outputs are possible: the
differential cross section $(d\sigma/dE_R)_T$ defined in
Eq.~(\ref{eq:dsigma_der}), the differential rate $(dR/dE_R)_T$ defined
in Eq.~(\ref{eq:dr_der_piecewise}) and the integrated expected rate
$S^{(0,1)}_{[E_1^{\prime},E_2^{\prime}]}$ given in
Eqs.~(\ref{eq:s0}-\ref{eq:sm}).}
\label{fig:code_structure}
\end{figure}

WimPyDD is a modular, customizable and flexible code written in Python
3.7 that allows to implement the formulas outlined in
Section~\ref{sec:theory} with a full control on each input.  The
versatility of the code is at the expense of some degree of
complexity. However most functionalities can be used with default
choices, so the beginner can start producing sensible output with just
a few simple calls. The present introduction to the code will mostly
describe default features, with a few mentions to possible
generalizations. 
More details about WimPyDD's functionalities and example files are available at:

\begin{Verbatim}[frame=single,xleftmargin=1cm,xrightmargin=1cm,commandchars=\\\{\}]
  https://wimpydd.hepforge.org
\end{Verbatim}

\noindent where also the code and additional explanations are
available. Also use the {\verb help() } function from the Python command line
to get detailed information about any component of the code.

The structure of the code is summarized in
Fig.~\ref{fig:code_structure}. In such figure diamond--shaped elements
indicate the input from the user, oval shapes contain classes and
rectangles represent functions. Three final outputs are possible, the
differential cross section $(d\sigma/dE_R)_T$ defined in
Eq.~(\ref{eq:dsigma_der}), the differential rate $(dR/dE_R)_T$ defined
in Eq.~(\ref{eq:dr_der_piecewise}) and the integrated expected rate
$S^{(0,1)}_{[E_1^{\prime},E_2^{\prime}]}$ given in
Eq.~(\ref{eq:rate_rbar_inelastic}), that includes the detector's response:
 the calculation of
$(d\sigma/dE_R)_T$ requires the minimal set of inputs, the nuclear
target(s) $T$ and an effective Hamiltonian ${\cal H}$; in addition to
them, to evaluate $(dR/dE_R)_T$ the halo function $\eta(v)$ is
required; finally the evaluation of the integrated rate
$R_{[E_1^{\prime},E_2^{\prime}]}$ requires to fix a specific
experimental set--up. A simple routine that calculates exclusion plots
in the familiar mass--cross section plane is also provided.

\subsection{Installation}
\label{sec:quick_start}

To install the latest version of WimPyDD type:

\begin{Verbatim}[frame=single,xleftmargin=1cm,xrightmargin=1cm,commandchars=\\\{\}]
git clone https://phab.hepforge.org/source/WimPyDD.git
\end{Verbatim}

\noindent in the working directory, or download the code from
WimPyDD's homepage. This will create a folder WimPyDD.  All the
libraries required by the code are standard, such as matplotlib,
numpy, scipy and pickle. To import WimPyDD type something like:

\begin{Verbatim}[frame=single,xleftmargin=1cm,xrightmargin=1cm,commandchars=\\\{\}]
import WimPyDD as WD
\end{Verbatim}

\subsection{Signal routines}
\label{sec:signal_routines}

To calculate the differential cross section $(d\sigma/dE_R)_T$ (in
cm$^2$/keV) use:

\begin{Verbatim}[frame=single,xleftmargin=1cm,xrightmargin=1cm,commandchars=\\\{\}]
  WD.dsigma_der(element_obj,hamiltonian_obj,mchi,v,\\
  er,**args)
\end{Verbatim}

\noindent with {\verb mchi } the WIMP mass $m_\chi$ in GeV,
{\verb v } the WIMP incoming speed $v$ in km/sec,
{\verb er } the recoil energy in keV, and where {\verb element_obj }
defines a nuclear target
$T$ (see Section~\ref{sec:element}) while {\verb hamiltonian_obj }
defines an effective Hamiltonian
(see Section~\ref{sec:eft_hamiltonian}). 

To evaluate the differential rate $dR/dE_R$ 
use the routine:

\begin{Verbatim}[frame=single,xleftmargin=1cm,xrightmargin=1cm,commandchars=\\\{\}]
  WD.diff_rate(target_obj,hamiltonian_obj,mchi,\\
  energy,vmin,delta_eta, **args)
\end{Verbatim}

\noindent {\verb target_obj } is either a single element or a
combination of elements with stoichiometric coefficients (see Section~\ref{sec:target}). The output of {\verb diff_rate } is in events/kg/day/keV
if the {\verb exposure } argument is not passed (its default value is 1 kg day).
The argument {\verb energy } is in keV and is interpreted as
the electron-equivalent energy $E_{ee}$ for the elements that have the
{\verb quenching } attribute, otherwise as the recoil energy
$E_R$ (see Eq.~(\ref{eq:start2}) and below for the definitions of $E_{ee}$, $E_R$ and of the quenching factor).
 Quenching is an attribute of the experimental set-up, so
can be added to a target using the {\verb experiment } class
(see Section~\ref{sec:element} and, for a specific example,
Section~\ref{sec:example_diff_rate}).

Finally, {\verb vmin } and {\verb delta_eta } are two arrays
  containing the $v_k$ and $\delta\eta_k^{(0,1)}$ of
  Eq.~(\ref{eq:delta_eta01_piecewise}) in km/s and (km/s)$^{-1}$, respectively
  (WimPyDD provides the routine {\verb streamed_halo_function } to calculate them,
  see Appendix~\ref{sec:halo}).
    
In order to calculate the integrated rate of Eq.~(\ref{eq:r_eta}) including the detector's response,
 use: 

\begin{Verbatim}[frame=single,xleftmargin=1cm,xrightmargin=1cm,commandchars=\\\{\}]
  WD.wimp_dd_rate(exp_obj, hamiltonian_obg, vmin,\\
  delta_eta, mchi,**args)
\end{Verbatim}

\noindent where {\verb exp_obj } is instantiated by the {\verb experiment } class
and contains all the required information about the experimental
set-up, including the energy bins where the rate is integrated
(see Appendix~\ref{app:exp}).  The routine returns two arrays with the centers
of the bins and the corresponding expected rates calculated using
Eq.~(\ref{eq:rate_rbar_inelastic}).

In all WimPyDD routines by default $\delta$=0 and $j_\chi$=1/2; such values
can be changed using the arguments {\verb delta } and {\verb j_chi }.
Moreover in all signal routines the parameters entering the Wilson coefficients of the
effective Hamiltonian
are passed through the keyworded arguments list {\verb **args }.
Additional default-valued arguments allow to tune the output and behavior
of all WimPyDD routines, type {\verb help } from the Python command line to
learn more. Explicit examples of the use of the signal routines are provided in
Section~\ref{sec:examples}.

\subsection{Main classes}
\label{sec:classes}

\subsubsection{Element}
\label{sec:element}

{\bf Main attributes:} {\verb a } (array with isotopic atomic numbers),
{\verb abundance } (array of isotopic natural abundances),
{\verb quenching } (quenching factor - only present if the element belongs
to the target of an {\verb experiment } object),
{\verb func_w } (array of isotopic nuclear response functions $W^{\tau\tau^{\prime}}_l$),
{\verb isotopes } (array with list of isotopes), {\verb mass } (array with isotope masses),
{\verb name } (full name of element), {\verb nt_kg }
(array with isotopic numbers of targets per kg) , {\verb spin } (array with isotope spins),
{\verb symbol } (element symbol),  {\verb z } (atomic number). 

WimPyDD provides a list of 20 pre-defined elemental targets that can be
shown with the command:
\begin{Verbatim}[frame=single,xleftmargin=1cm,xrightmargin=1cm,commandchars=\\\{\}]
WD.list_elements()\\
WD.Al WD.Ar WD.C WD.Ca WD.F WD.Fe\\
WD.Ge WD.H WD.He WD.I WD.Mg WD.N \\
WD.Na WD.Ne WD.Ni WD.O WD.S WD.Si\\
WD.W WD.Xe
\end{Verbatim}

\noindent To implement additional elements see Appendix~\ref{app:add_elements}.

\subsubsection{Target}
\label{sec:target}
{\bf Main attributes:} {\verb element } (array with list of elements), {\verb formula }
(molecular formula), {\verb mass } (total mass), {\verb n }
(array with stoichiometric coefficients), {\verb nt_kg }
(array with number of targets per kg for each element of the target).

A nuclear target can by any combination of {\verb element }
objects with stoihiometric coefficients. For instance, to define an
octafluoropropane ($C_3 F_8$) target:

\begin{Verbatim}[frame=single,xleftmargin=1cm,xrightmargin=1cm,commandchars=\\\{\}]
  c3f8=3*WD.C+8*WD.F
\end{Verbatim}

The output of any linear combination of {\verb element } and/or {\verb target } objects is a {\verb target } object. The same result is obtained by
directly invoking the {\verb target } class with a string argument
containing the element names and the stoichiometric coefficients,
i.e.:

\begin{Verbatim}[frame=single,xleftmargin=1cm,xrightmargin=1cm,commandchars=\\\{\}]
  c3f8=WD.target('C3F8')
\end{Verbatim}

\noindent It is at this stage that, in order to correctly parse the
input string, the element string names are required to begin with a
capital letter (i.e. {\verb WD.target('CaWO4') } is parsed correctly, while
{\verb WD.target('cawo4') } is not). The {\verb target } class needs also to be
invoked for a pure elemental target (the same result is obtained by
multiplying an {\verb element } object times 1):

\begin{Verbatim}[frame=single,xleftmargin=1cm,xrightmargin=1cm,commandchars=\\\{\}]
  germanium_target1=WD.target('Ge')\\
  germanium_target2=1*WD.Ge
\end{Verbatim}
  
\noindent Notice that inside a {\verb target } object
{\verb element } instantiations are sensitive to the environment. For
instance, the {\verb element.nt_kg } attribute contains the number of
target per unit mass $N_T$ for each element (see for instance Eq.~(\ref{eq:dr_der})).
This quantity is different if it refers to one kg of carbon or fluorine, or if it refers to 1 kg of $C_3F_8$:

\begin{Verbatim}[frame=single,xleftmargin=1cm,xrightmargin=1cm,commandchars=\\\{\}]
print(WD.F.nt_kg)\\
[3.16842105e+25]\\
print(c3f8.element[1])\\
fluorine, symbol F, atomic number 9, \\
average mass 17.689, 1 isotopes.\\
print(c3f8.element[1].nt_kg)\\
[2.56125255e+25]  
\end{Verbatim}

\subsubsection{Effective Hamiltonian} 
\label{sec:eft_hamiltonian}
{\bf Main attributes:} {\verb coeff_squared_list } (list of operators
products in scattering amplitude), {\verb couplings } (list of effective
operators), {\verb wilson_coefficients } (dictionary with effective operators and Wilson coefficients functions).

A specific effective Hamiltonian~Eq.(\ref{eq:H}) is set--up by a list of interaction
operators and their corresponding Wilson coefficients. In WimPyDD this
is done by instantiating the class {\verb eft_hamiltonian }, which
takes two arguments: a string containing a name of user's choice for
the Hamiltonian and a Python dictionary with the operators and the
Wilson coefficients:

\begin{Verbatim}[frame=single,xleftmargin=1cm,xrightmargin=1cm,commandchars=\\\{\}]
  wc=\{n1: func1 , n2: func2, ... \}  \\
  hamiltonian_obj=WD.eft_hamiltonian('name',wc)   
\end{Verbatim}

In WimPyDD two alternative sets of non--relativistic
effective operators can be used. The keys {\verb n1,n2,... }
of the dictionary {\verb wc } identify such operators and
are integer numbers for the base of operators introduced
in~\cite{haxton1,haxton2} or
tuples {\verb (X,s,l) } for the alternative base introduced
in~\cite{all_spins}.
Further information on the allowed choices for {\verb n1,n2,... }
is provided in Appendix~\ref{app:eft}.

In the same dictionary {\verb func1(par1,par2,...) },
{\verb func2(par1,par2,...) } are functions of arbitrary parameters
{\verb par1,par2,... } that can include $m_\chi$, $\delta$ and $q$.
They return two-dimensional arrays (or lists)
containing the Wilson coefficients $c_j^0,c_j^1$ in GeV$^{-2}$:

\begin{Verbatim}[frame=single,xleftmargin=1cm,xrightmargin=1cm,commandchars=\\\{\}]
def func1(w1,w2,...,mchi,...,delta,..,q):
    ...
   return [c0,c1]  
\end{Verbatim}

\subsubsection{Experiment}
\label{sec:load_response_function}
{\bf Main attributes:} {\verb data } (list of energy bins - can contain the
upper bounds used by {\verb mchi_vs_exclusion } routine),
{\verb efficiency } (efficiency function), {\verb exposure } (exposure  -
can be a list with a different exposure for each energy bin),
{\verb name } (name of experiment -
corresponds to the name of the folder with the experimental information), {\verb resolution } (energy resolution function),
{\verb response_functions } (dictionary with tables of integrated response functions),
{\verb target } (experimental target).

In order to include the response of the detector the {\verb wimp_dd_rate } routine requires the integrated response functions
$\left [\bar{{\cal R}}^a_{T, [E_1^{\prime},E_2^{\prime}]}\right
]_{jk}^{\tau\tau^{\prime}}$ of
Eq.~(\ref{eq:rbar}), whose calculation needs
information on the experimental set--up: the target, the energy bins,
the energy resolution, the efficiency, the quenching factors,
the exposure. All such information can be collected in a single Python object by
using the class {\verb experiment }:

\begin{Verbatim}[frame=single,xleftmargin=1cm,xrightmargin=1cm,commandchars=\\\{\}]
exp_obj=WD.experiment(name='exp')
\end{Verbatim}

\noindent that loads it from the folder  {\verb WimPyDD/Experiments/exp }
in the user's working directory (a detailed explanation on how
to initialize such directory is provided in Appendix~\ref{app:exp}).

For the effective Hamiltonian
{\verb hamiltonian_obj }, the experimental set up
{\verb experiment_obj } and WIMP spin {\verb j_chi } the
response functions are stored in the dictionary entry:

\begin{Verbatim}[frame=single,xleftmargin=1cm,xrightmargin=1cm,commandchars=\\\{\}]
  exp_obj.response_functions=\\
  \{(hamiltonian_obj,j_chi):r\} 
\end{Verbatim}

\noindent 
by the instruction: 

\begin{Verbatim}[frame=single,xleftmargin=1cm,xrightmargin=1cm,commandchars=\\\{\}]
  WD.load_response_functions(experiment_obj,
  hamiltonian_obj, j_chi)
\end{Verbatim}

\noindent  The values {\verb r } of such dictionary are tuples 
containing the tables of the response functions. The same tables
are also saved in the folder:

\begin{Verbatim}[frame=single,xleftmargin=1cm,xrightmargin=1cm,commandchars=\\\{\}]
WimPyDD/Experiments/exp/Response_functions/  
\end{Verbatim}

\noindent (see Section~\ref{sec:handling} for more details).
 If such folder already contains the tabulated response functions
{\verb load_response_functions } skips the calculation and directly loads them. The function
{\verb wimp_dd_rate } automatically
calls {\verb load_response_functions } if
the required response functions are not in the {\verb exp_obj.response_functions }
dictionary.

\subsection{Handling the response functions set}
\label{sec:handling}

An important feature of WimPyDD is that for a
given experimental set-up it allows to use the same pool of response
functions to study the WIMP parameters space of different effective
Hamiltonians. In particular the dependence of signals on a specific theoretical
model is factorized in the Wilson coefficients $c_j^{\tau}$, while for
all possible combinations $c_j^{\tau}c_k^{\tau^{\prime}}$ the
remaining information (nuclear physics + experimental) is stored in
the response functions $[\bar{{\cal
      R}}^a_{T, [E_1^{\prime},E_2^{\prime}]}(E_R)]^{\tau\tau^{\prime}}_{jk}$.
This means that the calculation of signals for different
Hamiltonians instantiated by {\verb WD.eft_hamiltonian }  can share the same set of response
functions. Whether this is done or not can be decided by the user, and
to do it properly requires to know a few more details on how WimPyDD
handles them.

For a specific Wilson coefficients dictionary {\verb wc } the
{\verb WD.eft_hamiltonian  } class
stores in the attribute {\verb coeff_squared_list } a lists of
all the possible contributions to the squared amplitude, including interferences,
which is then used by {\verb WD.load_response_function } to save
or load the corresponding response functions. In particular, the names of
the files where the response functions are stored is
constructed from the keys of {\verb wc }.

For instance, if the two following Hamiltonians {\verb h1 } and {\verb h2 }
are defined, with the functions
{\verb func1(w1,w2) },  {\verb func2(w3,w4) }, {\verb func3(w5,w6) }
of arbitrary parameters {\verb w1,w2,w3,w4,w5,w6 } parameterizing their sets of
Wilson coefficients:

\begin{Verbatim}[frame=single,xleftmargin=0cm,xrightmargin=0cm,commandchars=\\\{\}]
  h1=WD.eft_hamiltonian('h1',\{1: func1, 3: func2\})\\
  h2=WD.eft_hamiltonian('h2',\{1: func3\}) \\
  print(h1)\\
  Hamiltonian name:h1\\
  Hamiltonian:func1(w1, w2)* O_1+func2(w3, w4)* O_3\\
  Squared amplitude contributions:\\
  O_1*O_1, O_3*O_1, O_3*O_3\\
  print(h2)\\
  Hamiltonian name:h2\\
  Hamiltonian:func3(w5, w6)* O_1\\
  Squared amplitude contributions:\\
  O_1*O_1  
\end{Verbatim}

the instructions:
\begin{Verbatim}[frame=single,xleftmargin=0cm,xrightmargin=0cm,commandchars=\\\{\}]
xenon1t=WD.experiment('xenon1t')\\
WD.load_response_functions(xenon1t,h1, j_chi=0.5)
\end{Verbatim}
\noindent save (or load, if already existing) 
the following set of response functions:
\begin{Verbatim}[frame=single,xleftmargin=0cm,xrightmargin=0cm,commandchars=\\\{\}]
  c1_c1.npy\\
  c3_c1.npy\\
  c3_c3.npy
\end{Verbatim}
\noindent in the directory {\verb WimPyDD/Experiments/xenon1t/Response_functions/spin_1_2 },
while:
\begin{Verbatim}[frame=single,xleftmargin=0cm,xrightmargin=0cm,commandchars=\\\{\}]
WD.load_response_functions(xenon1t,h2, j_chi=0.5)
\end{Verbatim}
\noindent will load
\begin{Verbatim}[frame=single,xleftmargin=0cm,xrightmargin=0cm,commandchars=\\\{\}]
  c1_c1.npy
\end{Verbatim}
\noindent from the same directory. So {\verb h1 } and
{\verb h2 } will share the same set of
response functions contained in {\verb c1_c1.npy }, because they both contain the
$\calO_1$ operator. This can optimize the calculations saving computing time
but must be handled carefully in the case of Wilson coefficients with an
explicit momentum dependence. In fact,
suppose that the third effective Hamiltonian {\verb h3 } is defined:
\begin{Verbatim}[frame=single,xleftmargin=0cm,xrightmargin=0cm,commandchars=\\\{\}]
  h3=WD.eft_hamiltonian('h3',\{(3,'q_dep'): func4\})\\
  print(h3)\\
  Hamiltonian name:h3\\
  Hamiltonian:c_3_q_dep(w7, q)* O_3_q_dep'(q)\\
  Squared amplitude contributions:\\
  O_3_q_dep'(q)*O_3_q_dep'(q)
\end{Verbatim}
\noindent with {\verb func4(w7,q) } a function that depends
explicitly on the transferred momentum $q$. In this case even if
{\verb h1 } and {\verb h3 } share the same operator $\calO_3$
they do not share the same set of integrated response functions,
because the explicit dependence on $q$ of the Wilson coefficient in
{\verb h3 } implies a different energy dependence in the integrals
of Eqs.~(\ref{eq:rbar}). The way WimPyDD handles a Hamiltonian term of
the type $\sum_{\tau} c_n^{\tau}(q) \mathcal{O}_{n}({\bf{q}}) \, t^{\tau}$
is to factorize the Wilson coefficient at
some momentum scale $q_0$ (see Eq.~(\ref{eq:r_factorization}))
and treat the the remaining part
$\calO_n^{\prime}\equiv c_n^{\tau}(q)/c^{\tau}_n(q_0)\calO_n$ as an operator different
from $\calO_n$.  Crucially, by adding an explanatory
string {\verb string } to the the keys of the Wilson coefficient dictionary the
user can have full control on which files the response
functions of $\calO_n$ and $\calO_n^{\prime}$ are saved in.
In particular when the base of operators of Ref.~\cite{haxton1,haxton2} is used
the integer {\verb n } can be substituted by the tuple {\verb (n,string) },
while in the case of the operator base of Ref.~\cite{all_spins}
the tuple  {\verb (X,s,l) } can be extended to  {\verb (X,s,l,string) }.
So in the example above the tuple
{\verb (3,'q_dep') } is used instead of the simple integer {\verb 3 }, so that the corresponding response functions are saved in
the file:
\begin{Verbatim}[frame=single,xleftmargin=0cm,xrightmargin=0cm,commandchars=\\\{\}]
  c3_q_dep_c3_q_dep.npy
\end{Verbatim}
\noindent and are not confused with those of {\verb h1 }.
Combinations such as:

\begin{Verbatim}[frame=single,xleftmargin=0cm,xrightmargin=0cm,commandchars=\\\{\}]
  ('Sigma',1,1,'meson_propagator')\\
  (5,'q_2_6_matching')
\end{Verbatim}
are legal options for the
coupling keys in the dictionary, leading to the following names of the
output files where the tables of the integrated response functions are saved:
\begin{Verbatim}[frame=single,xleftmargin=0cm,xrightmargin=0cm,commandchars=\\\{\}]
  c_Sigma_1_1_meson_propagator_c_Sigma_1_1_meson_propagator.npy\\
  c_5_q_2_6_matching_c_5_q_2_6_matching.npy
\end{Verbatim}

Interferences among different operators $\calO^{\prime}_n$ and $\calO_n$ originated from the same $n$
are automatically included and the corresponding response functions
are calculated and saved, or loaded: 
\begin{Verbatim}[frame=single,xleftmargin=0cm,xrightmargin=0cm,commandchars=\\\{\}]
  h4=WD.eft_hamiltonian('h4',\{1: lambda: [1,1],\\
  (1,'q_dep'): lambda q : [1/q,1/q]\})\\
  print(h4)\\
  Hamiltonian name:4\\
  Hamiltonian:c_1()* O_1+c_1_q_dep(q)* O_1_q_dep'(q)\\
  Squared amplitude contributions:\\
  O_1*O_1, O_1*O_1_q_dep'(q), O_1_q_dep'(q)*O_1,\\
  O_1_q_dep'(q)*O_1_q_dep'(q)
\end{Verbatim}

In summary, WimPyDD allows to share the same set of response functions
for different theoretical scenarios, but it is up to the user to do
that appropriately by carefully managing the key names in the Wilson
coefficient dictionaries of each {\verb WD.eft_hamiltonian } object.

\subsection{Examples}
\label{sec:examples}
\noindent In the present Section we introduce the potentialities of
WimPyDD by showing some explicit examples.

\subsubsection{Differential rate}
\label{sec:example_diff_rate}

\begin{figure}
\begin{center}
  \includegraphics[width=0.7\columnwidth]{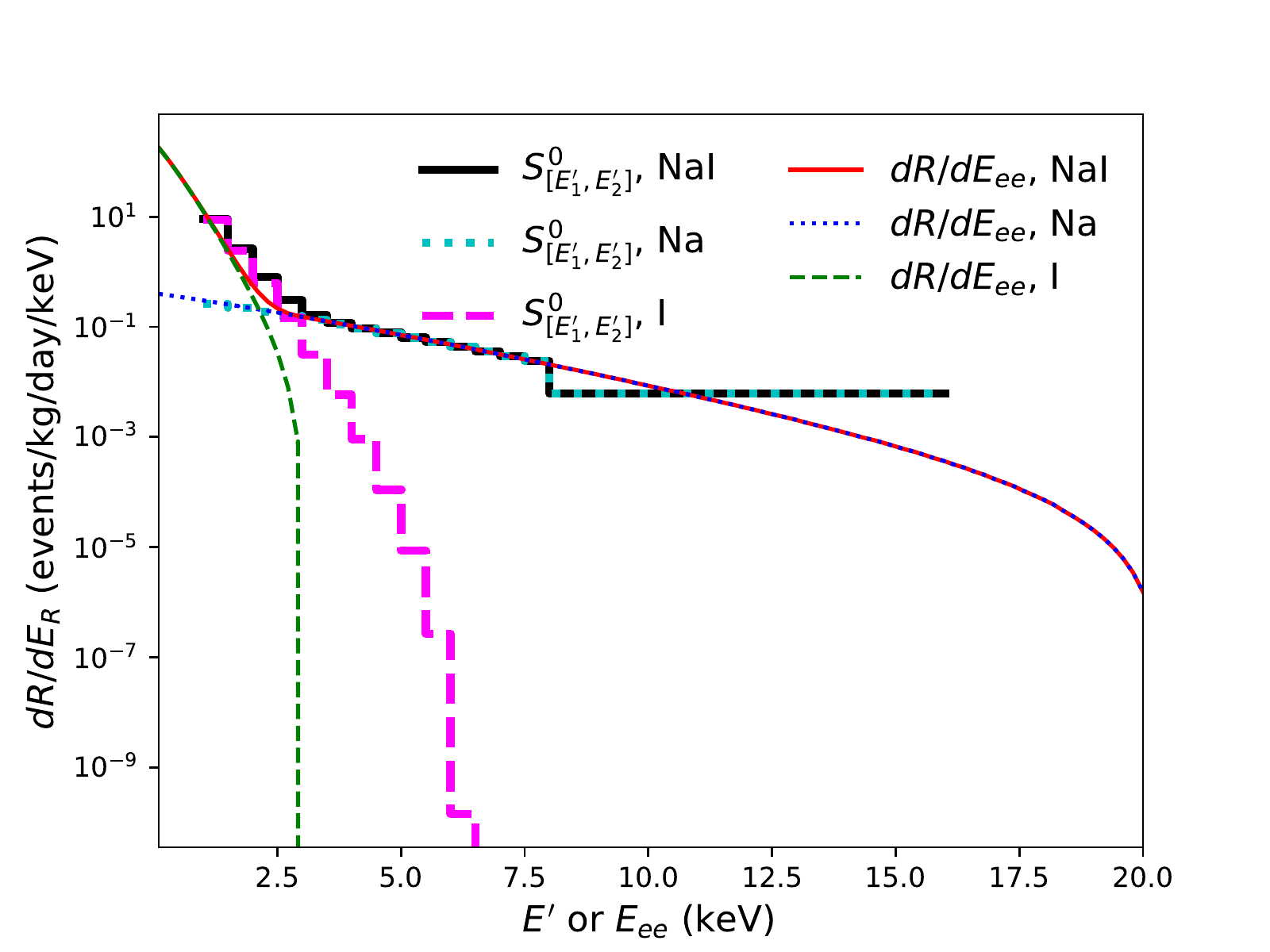}
  \end{center}
\cprotect\caption{Expected signals for the effective
  Hamiltonian of~Eq.(\ref{eq:hamiltonian_example1}) with $m_\chi$=20 GeV and $r$=1
  on sodium iodide.
  {\bf Thin lines:} output of {\verb diff_rate }
  for the differential rate $dR/dE_{ee}$ (Eq.~(\ref{eq:dr_der_piecewise}));
  {\bf thick lines:} output of the {\verb wimp_dd_rate }
  routine for the integrated rate $S^{(0)}_{[E_1^{\prime},E_2^{\prime}]}$
  (Eq.~(\ref{eq:rate_rbar_inelastic}))
  assuming binning and detector response of the DAMA experiment (the last bin is arbitrarily
  averaged over a wider range).
  {\bf Solid lines:} $NaI$; {\bf dotted lines:} contribution from scatterings off $Na$;
  {\bf dashed line:} contribution from scattering off $I$.  }
\label{fig:diff_rate_example}
\end{figure}

For this example let's set up the following effective Hamiltonian:

\begin{equation}
{\cal H}=c_1^0{\cal O}_1 t^0+c_1^1{\cal O}_1
t^1= \frac{1}{M^2}[(1+r){\cal O}_1 t^0+(1-r){\cal O}_1 t^1],
\label{eq:hamiltonian_example1}
\end{equation}

\noindent that depends on the two parameters $M$ and
$r$, that parameterize the WIMP--proton coupling $c_1^p=1/M^2$ and the
ratio between the WIMP--proton and the WIMP-neutron coupling
$r=c_n/c_p$. This is implemented by:

\begin{Verbatim}[frame=single,xleftmargin=1cm,xrightmargin=1cm,commandchars=\\\{\}]
  wc=\{1: lambda M, r=1 : [(1.+r)/M**2, (1-r)/M**2] \}  \\
  SI=WD.eft_hamiltonian('SI',wc)  \\ 
\end{Verbatim}

\noindent In the instruction above the function of the Wilson coefficient is initialized on the fly using the python {\verb lambda } built--in command. Any argument of the Wilson coefficients functions
can be assigned a default value, in the example above $r$=1~\footnote{The same
  variable can appear in more than one Wilson coefficient. Assigning it
  different default values in
    different Wilson coefficients may lead to unexpected results.}.
According to Table~\ref{tab:Haxton_operators} the operators $\mathcal{O}_1$ and
$ \mathcal{O}_{M,0,0}$ are equivalent. This means that the Hamiltonian defined above
can also be initialized setting:

\begin{Verbatim}[frame=single,xleftmargin=1cm,xrightmargin=1cm,commandchars=\\\{\}]
  wc=\{('M',0,0): lambda M, r=1 : [(1.+r)/M**2, \\
  (1-r)/M**2] \}. 
\end{Verbatim}

The calculation of the differential WIMP-nucleus scattering rate of
Eq.~(\ref{eq:dr_der}) requires a sodium-iodide target.  In a realistic
experiment like DAMA~\cite{dama_2018}, 
COSINE~\cite{cosine_modulation_2019} or ANAIS~\cite{anais_modulation_2019} sodium and iodine have different
quenching factors that need to be taken into account in order to obtain the correct spectral shape.
Notice, however, that the quenching is a property of both the
target and a specific detector. This means that by defining sodium
iodide as:

\begin{Verbatim}[frame=single,xleftmargin=1cm,xrightmargin=1cm,commandchars=\\\{\}]
nai_no_quenching=WD.Na+WD.I
\end{Verbatim}

\noindent the target will have no quenching attribute:

\begin{Verbatim}[frame=single,xleftmargin=1cm,xrightmargin=1cm,commandchars=\\\{\}]
  ['quenching' in dir(e) for e in nai_no_quenching.element]\\
  [False, False]
\end{Verbatim}

\noindent because in such object the experimental information is missing. 
If, however, the directory {\verb WimPyDD/Experiments/dama } contains
files with quenching information (see Appendix~\ref{app:exp} and
Table~\ref{tab:exp_file_list}) the command:

\begin{Verbatim}[frame=single,xleftmargin=1cm,xrightmargin=1cm,commandchars=\\\{\}]
dama=WD.experiment('dama')\\  
\end{Verbatim}

\noindent creates an {\verb experiment } object with a {\verb target }
attribute that now has quenching factors:
\begin{Verbatim}[frame=single,xleftmargin=1cm,xrightmargin=1cm,commandchars=\\\{\}]
  nai=dama.target
  ['quenching' in dir(e) for e in nai.element]
  [True, True]
\end{Verbatim}

\noindent So, as explained in Section~\ref{sec:signal_routines}, in order to calculate the differential rate 
as a function of the electron--equivalent energy one needs to use the target {\verb nai } instead of {\verb nai_no_quenching }  
in the {\verb diff_rate } function, so that the argument {\verb energy } 
 is interpreted as $E_{ee}$. As far as the halo function stream contributions
$\delta\eta_k^{(0)}$ are concerned they are calculated for a standard
Maxwellian with standard parameters by:
\begin{Verbatim}[frame=single,xleftmargin=1cm,xrightmargin=1cm,commandchars=\\\{\}]
  vmin,delta_eta0=WD.streamed_halo_function()\\
\end{Verbatim}

\noindent (see Appendix~\ref{sec:halo}). By default 1000 values are tabulated,
which is usually more than enough to accurately approximate the
velocity integral. Finally, the the differential rate is given by:

\begin{Verbatim}[frame=single,xleftmargin=1cm,xrightmargin=1cm,commandchars=\\\{\}]
  er_vec=np.linspace(0.1,20,100) \\
  mchi=20. \\
  diff_rate=[WD.diff_rate(nai, SI, mchi, er, \\
    vmin, delta_eta0,M=1e3) for er in er_vec] \\
  import matplotlib.pyplot as pl\\
  pl.plot(er_vec,diff_rate)
\end{Verbatim}
  \noindent in terms of the model parameters {\verb M } and {\verb r } that are passed as a keyworded, variable-length argument list.
  If some parameters are not passed
their default values are used if existing, otherwise they are fixed to
1.
The last command uses {\verb Matplotlib } to plot the output shown by
thin solid red curve in Fig.~\ref{fig:diff_rate_example}.
As already pointed out in this case the {\verb energy }
parameter is interpreted as the electron--equivalent energy $E_{ee}=q_T(E_R)E_R$
(see Appendix~\ref{app:response_functions}), since both sodium and iodine have
the {\verb quenching } attribute.

The behaviour of every WimPyDD routine can be tuned by changing
the default values of additional parameters that can be inspected issuingIn
a {\verb help } command. For instance the argument {\verb elements_list }
allows to plot the separate contributions of sodium and iodine:

\begin{Verbatim}[frame=single,xleftmargin=1cm,xrightmargin=1cm,commandchars=\\\{\}]
  na,i=nai.element
  diff_rate_na=[WD.diff_rate(nai, SI, mchi, er, \\
    vmin, delta_eta0,elements_list=[na],M=1e3) \\
    for er in er_vec] \\
  diff_rate_i=[WD.diff_rate(nai, SI, mchi, er, \\
    vmin, delta_eta0,elements_list=[i],M=1e3)\\
  for er in er_vec] \\
\end{Verbatim}

\noindent Such contributions are shown in Fig.~\ref{fig:diff_rate_example}
by the blue dotted and green dashed lines, respectively.

\subsubsection{Integrated rate with detector response}
The {\verb diff_rate } routine does not include the effects of experimental
binning and of the the response of the detector.
In order to include such effects the {\verb wimp_dd_rate } must be used, which requires to
instantiate an experimental set up with the {\verb experiment } class and to calculate the 
integrated response functions of Eq.~(\ref{eq:rbar}).
In particular, the following instructions:

\begin{Verbatim}[frame=single,xleftmargin=1cm,xrightmargin=1cm,commandchars=\\\{\}]
  WD.load_response_functions(dama,SI,j_chi=0.5)\\
  e,s0=WD.wimp_dd_rate(dama, SI, vmin, delta_eta0,\\
 mchi=20,M=1e3, r=1)
 e,s0_na=WD.wimp_dd_rate(dama, SI, vmin, delta_eta0,\\
 mchi=20,elements_list=[na],M=1e3, r=1)
 e,s0_i=WD.wimp_dd_rate(dama, SI, vmin, delta_eta0,\\
 mchi=20,elements_list=[i],M=1e3, r=1)
\end{Verbatim}

\noindent use the {\verb dama } object instantiated in Sec.~\ref{sec:example_diff_rate}
to produce as an output the thick lines in Fig.~\ref{fig:diff_rate_example}
according to the information about the binning and the detector response stored
in the {\verb WimPyDD/Experiments/dama/ } folder. In the same directory the file
  {\verb exposure.tab } contains an array with the exposures in each energy
  bin. In order to get the {\verb wimp_dd_rate } output in
  events/kg/day/keV such array must contain the inverses of the bin widths
  {\verb delta_e_1 } , {\verb delta_e_2 },...{\verb delta_e_n }  i.e. must have the
  form {\verb [1./delta_e_1 }, {\verb 1./delta_e_2 }...{\verb 1./delta_e_n]  } (see Appendix~\ref{app:exposure}).

On the other hand, the following instructions:
\begin{Verbatim}[frame=single,xleftmargin=1cm,xrightmargin=1cm,commandchars=\\\{\}]
def c9_tau(mchi,cross_section,r):\\
    hbarc2=0.389e-27\\
    mn=0.931\\
    mu=mchi*mn/(mchi+mn)\\
    cp=(np.pi*cross_section/(mu**2*hbarc2))**(1./2.)\\
    return cp*np.array([1.+r,1.-r])\\
    \\
c9=WD.eft_hamiltonian('c9',\{9: c9_tau\})\\
vmin,delta_eta1=WD.streamed_halo_function(\\
yearly_modulation=True,n_vmin_bin=1000)\\  
WD.load_response_functions(dama,c9)\\
e,s1=WD.wimp_dd_rate(dama, c9, vmin, delta_eta1,\\
cross_section=8.29e-33, mchi=9.3 , r=4.36)
\end{Verbatim}

\begin{figure}
\begin{center}
  \includegraphics[width=0.7\columnwidth]{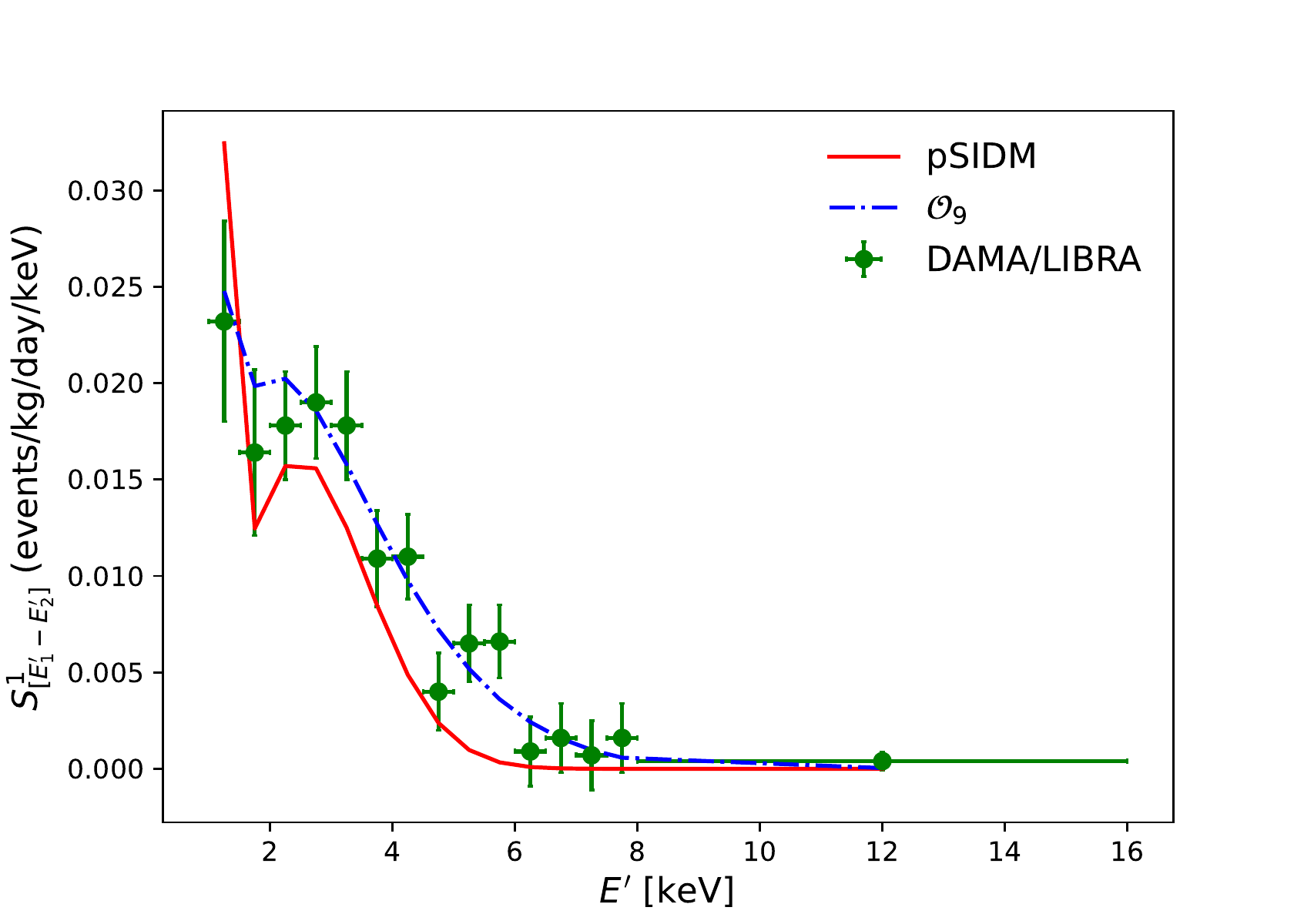}
  \end{center}
\caption{{\bf Error bars:} annual modulation amplitudes measured by the
  DAMA/LIBRA-Phase2 experiment~\cite{dama_2018} as a function of the
  observed recoil energy. {\bf Blue dotted line:} corresponding
  prediction if the WIMP--nucleus scattering is driven by the
  effective Hamiltonian ${\bf\mathcal{H}}_9= \sum_{\tau=0,1}
  c_9^{\tau}(m_\chi,\sigma_p,r) \mathcal{O}_{9}({\bf{r}}) \, t^{\tau}$
  for the best--fit of Ref.~\cite{dama_2018_sogang}. {\bf Red solid line:} the same quantity for the inelastic effective
  model driven by the effective Hamiltonian ${\bf\mathcal{H}}=
  \sum_{n=4,5,6}\sum_{\tau=0,1} c_n^{\tau} \mathcal{O}_{n}({\bf{r}})
  \, t^{\tau}$ in Table II of Ref.~\cite{dama_inelastic_eft_sogang}
  ($j_\chi$=1/2).}
\label{fig:dama_fit_examples}
\end{figure}

\noindent use the same {\verb dama } object to perform the following steps:

\begin{itemize}
\item define the function {\verb c9_tau } that returns the two couplings (in GeV$^{-2}$) $c_9^{\tau}$=$(c_9^{0},c_9^{1})$
calculated in terms of the three parameters $m_\chi$,
$r=c_9^n/c_9^p=(c_9^0-c_9^1)/(c_9^0+c_9^1)$ and
$\sigma_p=(c_9^p)^2\mu_{\chi N}^2/\pi(\hbar c)^2$ (in cm$^2$), i.e.
$c_9^p$ = $\sqrt{\pi \sigma_{ref}/(\mu_{\chi N}^2 (\hbar c)^2)}$,
$c_9^{\tau}$ = $c_9^p \times [1+r,1-r]$;
\item use the class {\verb eff_hamiltonian }  to define the model {\verb c9 }
  with effective Hamiltonian ${\bf\mathcal{H}}_9= \sum_{\tau=0,1} c_9^{\tau}(m_\chi,\sigma_p,r) \mathcal{O}_{9}({\bf{r}}) \, t^{\tau}$,
  initializing the Wilson coefficients with the
  dictionary {\verb {9:c9_tau} };   
\item calculate the modulated halo function $\delta\eta^1(v_k)$ of
  Eq.~(\ref{eq:delta_eta01_piecewise}) using the {\verb streamed_halo_function }
  routine with the argument  {\verb yearly_modulation=True } and saves the output in the
  two arrays {\verb vmin,delta_eta1 } ;
\item use the routine {\verb load_response_functions } to calculate or
  load (if already present in the directory {\verb WimPyDD/Experiments/dama })
  the response functions of Eq.~(\ref{eq:rbar}) for the {\verb c9 }
  effective model and a spin-1/2 WIMP (default value of spin);
\item use the routine {\verb wimp_dd_rate } to calculate the expected
  modulation amplitudes for the choice of parameters
  $\sigma_p=8.29\times 10^{-33}$ cm$^2$, $m_\chi$=9.3 GeV , r=4.36 (such values
  correspond to the best fit of the DAMA/LIBRA-Phase2
  experiment~\cite{dama_2018} for ${\bf\mathcal{H}_9}$ from
  Ref.~\cite{dama_2018_sogang}). The {\verb wimp_dd_rate } routine
  returns two arrays with the centers of the energy bins (in keV) and
  the corresponding predictions for the number of events.  The energy
  bins are taken from the {\verb data.tab } file in the
  {\verb WimPyDD/Experiments/dama } directory. 
\end{itemize}

\noindent The result is shown in Fig.~\ref{fig:dama_fit_examples} with the solid
  line and compared with the modulation amplitudes measured by
  DAMA/LIBRA-Phase2.

The same calculation can be easily extended to any
combination of effective couplings by adding entries in the dictionary
that parameterizes the Wilson coefficients.  For instance, in Table II
of Ref.~\cite{dama_inelastic_eft_sogang} the effective model
${\bf\mathcal{H}}= \sum_{n=4,5,6}\sum_{\tau=0,1} c_n^{\tau}
\mathcal{O}_{n}({\bf{r}}) \, t^{\tau}$ is discussed in terms of the eight 
parameters $m_\chi$, $\delta$ and ${\bf c}_0$ = [c$_4^0$,
  c$_4^1$, c$_5^0$, c$_5^1$, c$_6^0$, c$_6^1$]. The latter
couplings vector is further expressed in terms of a direction ${\bf
  \tilde{c}}_0$ = ${\bf c}_0$/ $|{\bf c}_0|$ in the coupling
space, and of an effective cross section parameterizing the couplings'
intensity, $\sigma_{eff}=|{\bf c}_0|^2\mu_{\chi N}^2/\pi(\hbar c)^2$.
In particular for $j_\chi$=1/2 the tension between the DAMA experiment and the
constraints from other experiments is found to be minimized by:
$m_\chi$=11.64 GeV, $\delta$ = 23.74 keV, ${\bf
  \tilde{c}}_0$ = [ -0.00139183,
  -0.00145925, -0.00318353, -0.01660634, 0.69199713, 0.72169939 ] and
$\sigma_{eff}$ =4.68$\times10^{-28}$ cm$^2$.  In WimPyDD such non--trivial effective
model can be easily implemented with a few lines of code:

\begin{Verbatim}[frame=single,xleftmargin=1cm,xrightmargin=1cm,commandchars=\\\{\}]
def c_tau(mchi,sigma,c_tilde):\\
    hbarc2=0.389e-27\\
    mn=0.931\\
    mu=mchi*mn/(mchi+mn)\\
    c_abs=(mu**2*hbarc2/(np.pi*sigma))**(-0.5)\\
    return c_abs*c_tilde\\
\\
wc=\{4: lambda mchi,s,c: c_tau(mchi,s,c)[0:2],\\
5: lambda mchi,s,c: c_tau(mchi,s,c)[2:4],\\
6: lambda mchi,s,c: c_tau(mchi,s,c)[4:6]\}\\
\\
c_456=WD.eft_hamiltonian('c_456',wc)\\

WD.load_response_functions(dama,c_456)
c_tilde=np.array([-0.00139183, -0.00145925,\\
-0.00318353, -0.01660634,  0.69199713, 0.72169939])\\
mchi=11.64102564102564\\
delta=23.73913043478261\\
cross_section=4.684124873205523e-28\\
pl.plot(*WD.wimp_dd_rate(dama,c_456,vmin,delta_eta1,\\
mchi,delta=delta,s=cross_section,c=c_tilde))
\end{Verbatim}

\noindent Notice that in this case the model is inelastic (a $\delta$
= $m_{\chi^{\prime}}$ - $m_{\chi}$ parameter different from zero is
used in Eq.~(\ref{eq:vmin}) for the $v_{min}$ parameter).  Moreover, as
already pointed out, in
the Wilson coefficients the {\verb mchi } and {\verb delta } arguments
are reserved and passed as positional arguments, while any other
parameter can acquire an arbitrary name and is passed as a keyworded
variable-length argument list. The corresponding result, plotted by
the last instruction, is shown in Fig~\ref{fig:dama_fit_examples} by the solid red line
and reproduces that of Fig.7 of~\cite{dama_inelastic_eft_sogang}.
Such prediction must be compared with the constraints from other
experiments. For instance, by filling up the directory
{\verb WimPyDD/Experiments/xenon1t } with all the relevant information for the
XENON1T experiment~\cite{xenon_2018} (see Appendix~\ref{app:exp}) one
can calculate the corresponding number of events in the range 3
PE$<S_1<$70 PE with the following instructions:

\begin{Verbatim}[frame=single,xleftmargin=1cm,xrightmargin=1cm,commandchars=\\\{\}]
xe=WD.target('Xe')\\
xenon1t=WD.experiment('xenon1t',xe)\\
WD.load_response_functions(xenon1t,c_456)\\
WD.wimp_dd_rate(xenon1t,c_456,vmin,delta_eta1,\\
mchi,delta=delta,s=cross_section,c=c_tilde)\\
(array([36.5]), array([4.78777365]))
\end{Verbatim}

In this case {\verb wimp_dd_rate } returns the center of the only ``bin'',
as well as a total number of 4.78 events for an exposure of 362440 kg day
(provided in the file {\verb exposure.tab }). Such value is indeed
consistent at 90\% C.L. with the 7 candidate events
measured in~\cite{xenon_2018}.  

The evaluation of the binned signals in this Section requires a triple integration,
over the WIMP velocity $v$, the nuclear recoil energy $E_R$ and the visible
energy $E^{\prime}$. WimPyDD performs the double integral
over $E_R$ and $E^{\prime}$ only once and stores the response functions
for later interpolation. Moreover the sums in
Eqs~(\ref{eq:delta_eta01_piecewise}) that substitute the
velocity integral are optimized for speed in {\verb numpy }. As a
consequence the {\verb wimp_dd_rate } routine is well suited for large
iterations over the input parameters {\verb par1,par2,... }. 

\subsubsection{Exclusion plot}
A familiar way to show constraints from DD experiments
is using exclusion plots in the mass--cross section plane. This
requires sometimes sophisticated procedure of background subtraction
and/or likelihood minimization. The {\verb WimPyDD } code provides the
simple routine {\verb mchi_vs_exclusion } that allows to estimate
the upper bound on any parameter $\lambda$ that can be factorized from
the rate by simply comparing the output of {\verb wimp_dd_rate } with
$\lambda$=1 to the experimental upper bounds on the count rate in each
energy bin (that the user can include in the {\verb data.tab }
experiment file) and taking the most stringent constraint.
This is obtained by calling {\verb mchi_vs_exclusion }
without passing the $\lambda$ argument (so that it is set to 1 by default)
or passing $\lambda$=1. For instance, in the case a
standard isoscalar Spin--Independent interaction driven by the
operator ${\cal O}_1$ the expected rate is proportional to the
WIMP--proton cross section $\sigma_p=(c_1^p)^2 \mu_{\chi N}^2/\pi$
(equal to the WIMP--neutron cross section $\sigma^n$).
To get the exclusion plot on $\sigma^p$ one must parameterize
the coupling $c_1^p$
in terms of $\sigma^p$ and call {\verb mchi_vs_exclusion }
without fixing $\sigma^p$ (or setting it to unity):

\begin{Verbatim}[frame=single,xleftmargin=0cm,xrightmargin=0cm,commandchars=\\\{\}]
hbarc2=0.389e-27\\
mn=0.931\\
wc_SI=\{1: lambda mchi,sigma_p: np.sqrt(sigma_p*np.pi/hbarc2)*\\
(1./mn+1./mchi)*np.array([2.,0])\}\\
SI=WD.eft_hamiltonian('SI',wc_SI)\\  
mchi,sigma=WD.mchi_vs_exclusion(xenon1t, SI, \\
vmin, delta_eta0)\\
pl.plot(mchi,sigma)
\end{Verbatim}

\begin{figure}
\begin{center}
  \includegraphics[width=0.49\columnwidth]{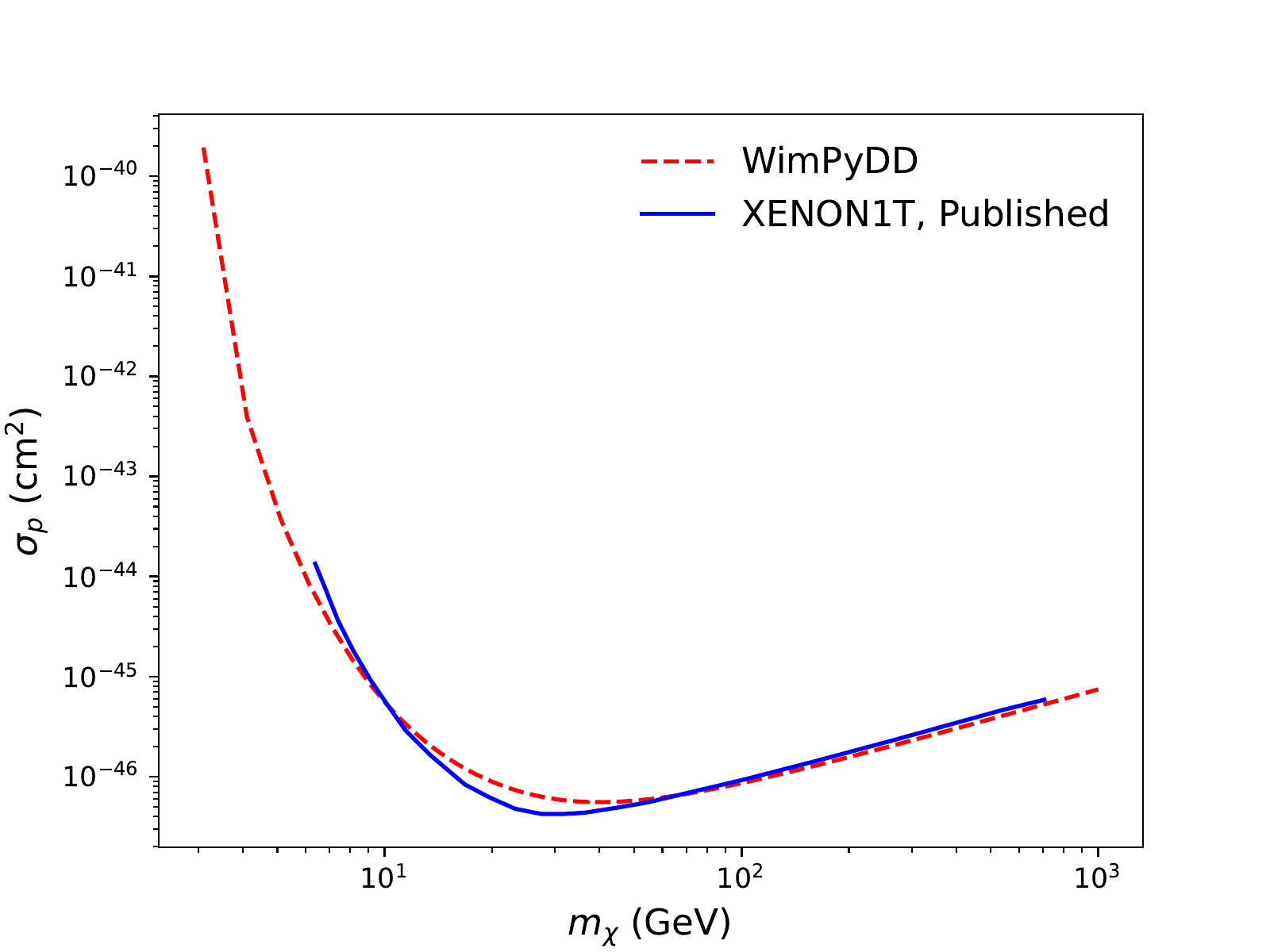}
  \includegraphics[width=0.49\columnwidth]{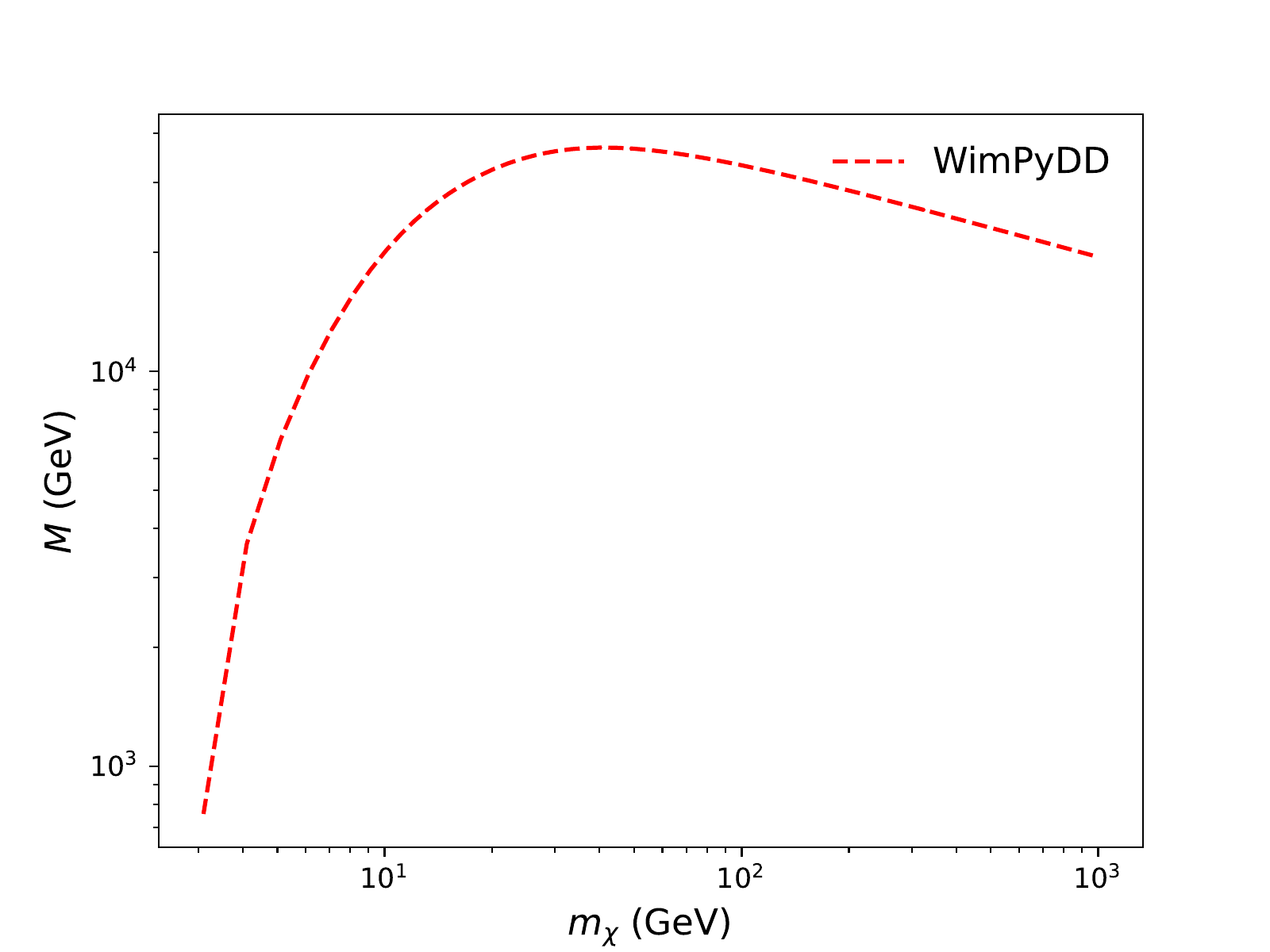}  
  \end{center}
\caption{XENON1T exclusion plot for a standard spin--independent interaction
  as a function of the WIMP mass $m_\chi$. {\bf Left--hand plot:} upper
  bound on WIMP--proton cross section; {\bf Right--hand plot:} upper bound
  on the effective scale $M$.}
\label{fig:xenon1t_exclusion}
\end{figure}

The output is shown in the left--hand plot of
Fig.~\ref{fig:xenon1t_exclusion} with the red dashes, while the solid blue
line is result of Ref.~\cite{xenon_2018}.

An appropriate reparameterization of the Wilson coefficients in
{\verb eft_hamiltonian } allows to use {\verb mchi_vs_exclusion } to get
exclusion plots on any parameter the expected rate is proportional to.
For instance, the following few lines get the {\it lower} bound on the
effective scale $M$ if the Wilson coefficient is parameterized as
$c_1^p=1/M^2$:

\begin{Verbatim}[frame=single,xleftmargin=0cm,xrightmargin=0cm,commandchars=\\\{\}]
SI_M=WD.eft_hamiltonian('SI_M',\{1: lambda M: \\
1./M**2*np.array([2,0])\})\\
mchi,one_over_M4=WD.mchi_vs_exclusion(xenon1t, SI_M, vmin,\\
delta_eta0)\\
pl.plot(mchi,one_over_M4**(-1./4.))\\
\end{Verbatim}

\noindent
In this case, since the expected rate is proportional to
$M^{-4}$, an upper bound on the latter quantity is obtained if
{\verb mchi_vs_exclusion } is called without the $M$ argument
(this is equivalent to passing {\verb M=1 } since {\verb M } has no default value). The plot
of the output in terms of $M$ = $(M^{-4})^{-1/4}$
is shown in the right--hand plot of
Fig.~\ref{fig:xenon1t_exclusion}.


\acknowledgments The research of I.J., S.K. and S.S was supported by the National
Research Foundation of Korea(NRF) funded by the Ministry of Education
through the Center for Quantum Space Time (CQUeST) with grant number
2020R1A6A1A03047877 and by the Ministry of Science and ICT with grant
number 2019R1F1A1052231. GT is supported by a TUM University
Foundation Fellowship and the Collaborative Research Center SFB1258.
\appendix

\section{Response functions}
\label{app:response_functions}
In this Section we derive explicitly Eq.~(\ref{eq:rbar}) and the
expressions of the integrated response functions discussed in
Section~\ref{sec:theory}.

The expected number of events in a WIMP DD experiment in
the interval of visible energy $E_1^{\prime}\le E^{\prime}\le
E_2^{\prime}$ is:

\begin{equation}
  R_{[E_1^{\prime},E_2^{\prime}]}(t)=\sum_T\int_{E_1^{\prime}}^{E_2^{\prime}}d E^{\prime}\;\left(\frac{dR}{d E^{\prime}} \right)_T(t)
  \label{eq:start1},
\end{equation}

\noindent where $T$ indicates the target nuclei present in the detector.
For a given target $T$: 

\begin{equation}
 \left(\frac{dR}{d E^{\prime}} \right)_T(t)=\int_0^\infty dE_R \left(\frac{dR}{d E_R}\right )_T(t) {\cal
   G}\left [E^{\prime},E_{ee}\right ]\epsilon(E^{\prime})
  \label{eq:start2}.
\end{equation}

In the equations above $E_R$ is the recoil energy deposited in the
scattering process (indicated in keVnr), while $E_{ee}$=$q_T(E_R)E_R$
(indicated in keV) is the fraction of $E_R$ that goes into
ionization and scintillation (wth $q_T(E_R)$ the target quenching
factor) and ${\cal G}(E^{\prime},E_{ee})$ is the effect of the energy
resolution (so that $E^{\prime}$ is measured instead of $E_{ee}$
through a calibration procedure\footnote{In the case of liquid scintillators such as XENON1T~\cite{xenon_2018} the electron--equivalent energy is substituted by the 
average number $<\nu>$ of  photo-electrons (PE), the visible energy is substituted 
by the observed number of PE $S_1$ and the light--yield  ${\cal L}$ takes the place of the quenching factor, with $S_1$ = ${\cal L} <\nu>$.}). Finally $\epsilon(E^{\prime})$ is
the measured experimental acceptance. The differential
rate is given by:

\begin{equation}
\left(\frac{dR}{dE_R}\right)_T(t)=M N_T T_0 \int_{v_{T,min}(E_R)}^{v_{esc}} \frac{\rho_{\chi}}{m_{\chi}}
v \left(\frac{d\sigma}{dE_R}\right)_T\, f(\vec{v},t) d^3v,
\label{eq:dr_der}
  \end{equation}

\noindent with $f(\vec{v},t)$ the WIMP velocity distribution in the
Earth's rest frame. In Eq.~\ref{eq:dr_der} $N_T=N_A 1000/m_{N}$
($N_A$=Avogadro number) is the number of targets in one kg of detector, $M$ is
the mass of the detector (so that $M N_T$ is the total number of
targets $T$), $T_0$ the time of exposure, $\rho_\chi$ is the mass
density of Dark Matter in the neighborhood of the Sun, $m_\chi$ the
WIMP mass (so that $\rho_\chi/m_\chi$ represents the WIMP number
density). In particular, using the piece--wise
definition of~(Eq.~\ref{eq:delta_eta_piecewise}) in Eq.~(\ref{eq:dr_der})
one gets:

\begin{eqnarray}
\left(\frac{dR}{dE_R}\right)_T(t)&=&M N_T T_0
\int_{v_{T,min}(E_R)}^{v_{esc}} \frac{\rho_{\chi}}{m_{\chi}} v
\left[\frac{d\sigma}{dE_R}(v)\right]_T\, \sum_k^{N_s} \lambda_k(t)
\delta(v-v_k) dv,\nonumber\\ &=& M N_T T_0
\sum_{v_k>v_{min}(E_R)}^{N_s} \frac{\rho_{\chi}}{m_{\chi}} v_k^2
\left[\frac{d\sigma}{dE_R}(v_k)\right]_T\delta\eta_k(t)\Theta(v_k-v),
\label{eq:dr_der_piecewise}  
\end{eqnarray}

Indicating with $j_T$ the target spin~\cite{haxton2}:

\begin{equation}
\left( \frac{d\sigma}{d E_R}\right )_T=\frac{2 m_T}{4\pi}\frac{c^2}{v^2}\left
     [\frac{1}{2j_{\chi}+1} \frac{1}{2j_{T}+1}\sum_{spin}|{\cal M}(q)|^2_T 
       \right ],
\label{eq:dsigma_der}
\end{equation}

\noindent is the differential cross section in terms of the scattering
amplitude averaged over the initial WIMP and target spins. In
non--relativistic effective theory the latter is given by:

\begin{equation}
  \frac{1}{2j_{\chi}+1} \frac{1}{2j_{T}+1}\sum_{spin}|{\cal M}(q)|^2_T=
  \frac{4\pi}{2j_T+1}\sum_{\tau\tau^{\prime}}\sum_l R_l^{\tau\tau^{\prime}} (q,v)W_{l,T}^{\tau\tau^{\prime}}(q).
\label{eq:haxton_40}
\end{equation}

\noindent In the equation above the WIMP and nuclear physics are
factorized in the two $R_l^{\tau\tau^{\prime}}$ and
$W_{l,T}^{\tau\tau^{\prime}}$ functions, respectively, with
$\tau,\tau^{\prime}=0,1$ the nuclear isospin while
$l$=$M$,$\Sigma^{\prime\prime}$
,$\Sigma^{\prime}$,$\Phi^{\prime\prime}$, $\Phi^{\prime\prime}M$,
$\tilde{\Phi}^{\prime}$,$\Delta$, $\Delta\Sigma^{\prime}$ represent
one of the possible nuclear interaction types, in the single--particle
interaction limit~\cite{haxton1,haxton2}. Moreover, in
Eqs.~(\ref{eq:dsigma_der}) and~(\ref{eq:haxton_40}) $q$ represents the
transferred momentum:

\begin{equation}
  q=\sqrt{2 m_T E_R}.
  \label{eq:q_E_R}
  \end{equation}

\noindent Explicit expressions of the WIMP response functions
$R_l^{\tau\tau^{\prime}}$ are provided in~\cite{haxton2}
and~\cite{all_spins} for the two sets of operators discussed
in Appendix~\ref{app:eft}.  They can be decomposed as:

\begin{equation}
R_l^{\tau\tau^{\prime}}(q,v)=R_{0,l}^{\tau\tau^{\prime}}(q)+R_{1,l}^{\tau\tau^{\prime}}(q)(v^2-v_{T,min}^2),
\label{eq:vel_dep}  
\end{equation}

\noindent while power/exponential expansions of the nuclear response
functions $W_{l,T}^{\tau\tau^{\prime}}$ have been provided in
\cite{haxton2} and \cite{catena} for most targets $T$ used in direct
detection experiments. A direct substitution of Eqs.~(\ref{eq:start2},
\ref{eq:dr_der}, \ref{eq:dsigma_der}, \ref{eq:haxton_40},
\ref{eq:vel_dep}) in Eq.~(\ref{eq:start1}) yields:

\begin{eqnarray}\nonumber
  &&R_{[E_1^{\prime},E_2^{\prime}]}(t)=\frac{\rho_\chi}{m_\chi}
  \frac{1}{\pi} \sum_T \int_{E_1^{\prime}}^{E_2^{\prime}}d E^{\prime}\; \int_0^\infty d E_R\; \int_{v_{T,min}(E_R)}^{\infty}d^3 v\; f(\vec{v},t)\frac{c^2}{v}\\\nonumber
  && \times N_T M T_0 \frac{m_T}{2}{\cal G}\left[E^{\prime},q_T(E_R)E_R \right ]\epsilon(E^{\prime})\\
  &&\times \frac{4\pi}{2j_T+1}\sum_{\tau\tau^{\prime}}\sum_l\left \{ R_{0,l}^{\tau\tau^{\prime}}(q)+R_{1,l}^{\tau\tau^{\prime}}(q)\left [v^2-v^2_{T,min}(E_R) \right ]\right \} W_{l,T}^{\tau\tau^{\prime}}(q).
  \label{eq:rate_before_inversion}
  \end{eqnarray}

\noindent Introducing the speed velocity distribution:

\begin{equation}
f(v,t)\equiv\int d\Omega_v f(\vec{v},t),\,\, v\equiv |\vec{v}|,
\end{equation}

\noindent the order of the two integrals in $E_R$ and $v$ can be
inverted~\cite{generalized_halo_indep}. In particular:

\begin{equation}
\int_0^\infty d E_R\; \int_{v_{T,min}(E_R)}^{\infty} d v\rightarrow \int_{v_{T^*}}^\infty dv\;\int_{E_R^{min}(v)}^{E_R^{max}(v)} d E_R
  \end{equation}

\noindent with $E_R^{min,max}(v)$ and $v_{T^*}$ given by
Eqs.~(\ref{eq:er_max_min}) and~(\ref{eq:vstar}). In this way
Eq.~(\ref{eq:rate_before_inversion}) can be recast in the form:

\begin{equation}
R_{[E_1^{\prime},E_2^{\prime}]}(t)=\int_{v_T^*}^\infty d v {\cal H}_{[E_1^{\prime},E_2^{\prime}]}(v) f(v,t), 
\end{equation}

\noindent with:

\begin{eqnarray}
  {\cal H}_{[E_1^{\prime},E_2^{\prime}]}(v)&=&\frac{\rho_\chi}{m_\chi} \frac{1}{\pi} \frac{c^2}{v}\sum_T
  \int_{E_R^{min}(v)}^{E_R^{max}(v)} d E_R\left\{ {\cal R}^0_{T}(E_R) +\right.
      \left .{\cal R}^1_{T}(E_R)\left(v^2-v^2_{T,min}(E_R) \right )\right\}\label{eq:Hv}\\
    {\cal R}^i_{T}(E_R)&=&\int_{E_1^{\prime}}^{E_2^{\prime}}d E^{\prime}\; r^i_{T}(E_R,E^\prime),\,\,\,i=0,1,  \label{eq:r_T_i}
\end{eqnarray}

\noindent and:

\begin{equation}
r^i_{T}(E_R,E^\prime)=M T_0 N_T\frac{m_T}{2}\epsilon(E^{\prime}){\cal
  G}[E^{\prime},q_T(E_R)E_R]\frac{4\pi}{2
  j_T+1}\sum_{\tau\tau^{\prime}}\sum_l
{\cal R}_{l,T}^{i,\tau\tau^{\prime}}(q)W_{l,T}^{\tau\tau^{\prime}}(q). \label{eq:differential_r}
\end{equation}

\noindent The speed distribution $f(v)$ can be written
as~\cite{generalized_halo_indep}:

\begin{equation}
  f(v,t)\equiv -v\frac{d}{dv}\eta(v,t),
  \label{eq:d_eta}
  \end{equation}
\noindent so that integrating by parts one gets Eq.~(\ref{eq:r_eta})
where:

\begin{equation}
{\cal R}_{[E_1^{\prime},E_2^{\prime}]}(v)=\frac{d}{dv}\left [ v {\cal H}_{[E_1^{\prime},E_2^{\prime}]}(v) \right ],
\label{eq:Rv}
\end{equation}

Substituting Eq.~(\ref{eq:delta_eta_piecewise}) in
Eq.~(\ref{eq:r_eta}), with ${\cal R}$ and ${\cal H}$ given by
Eqs.~(\ref{eq:Rv},\ref{eq:Hv}) one gets:

\begin{eqnarray}
&&R_{[E_1^{\prime},E_2^{\prime}]}(t)=\sum_T\frac{\rho_\chi}{m_\chi}  \frac{1}{\pi} 
 \int_{v_{T^*}}^{\infty} dv \frac{d}{dv}\left\{
 v \frac{c^2}{v}\int_{E_R^{min}(v)}^{E_R^{max}(v)}dE_R\left\{{\cal R}^0_{T,[E_1^{\prime},E_2^{\prime}]}(E_R)\right .\right .+\nonumber \\
 &&\left .\left .{\cal R}^1_{T,[E_1^{\prime},E_2^{\prime}]}(E_R)[v^2-v_{T,min}(E_R)^2]\right\}
  \right\}\times \sum_{k=1}^{N_s}
  \delta\eta_k(t)\theta(v_k-v)=\nonumber\\
&& \sum_T\frac{\rho_{\chi}}{m_{\chi}} \sum_{k=1}^{N_s} \delta\eta_k(t)\times \nonumber\\
  &&  \int_{E_R^{min}(v_k)}^{E_R^{max}(v_k)}dE_R\left\{{\cal R}^0_{T,[E_1^{\prime},E_2^{\prime}]}(E_R)+{\cal R}^1_{T,[E_1^{\prime},E_2^{\prime}]}(E_R)[v^2_k-v^2_{T,min}(E_R)]\right\},
  \label{eq:proof_r_dv_de}
\end{eqnarray}

\noindent which is equivalent to Eq.~(\ref{eq:r_dv_de}).

\section{Sets of non--relativistic effective operators}
\label{app:eft}
WimPyDD allows to use two alternative sets of non--relativistic
effective operators for the Hamiltonian of Eq.~(\ref{eq:H}): the set
introduced in Ref.~\cite{haxton1,haxton2} for a WIMP spin
$j_\chi\le$1/2 and the set of Ref.~\cite{all_spins}, valid for any
value of $j_\chi$.

In Refs.~~\cite{haxton1,haxton2} the operators $\mathcal{O}_{j}$ have
been indicated with progressive numbers from 1 to 16. However
$\mathcal{O}_2$ is neglected because it is quadratic in the WIMP incoming
velocity while $\mathcal{O}_{16}$ can be written as a linear
combination of $\mathcal{O}_{12}$ and $\mathcal{O}_{15}$.  WimPyDD allows to
use any of the remaining 14 independent operators by setting the keys of the
{\verb wc } dictionary to one of the integers 1,3,4,...,15.

In Ref.~\cite{all_spins} such base has been extended by the set of
irreducible operators $\mathcal{O}_{X,s,l}$ classified in terms of the
5 basic nuclear currents $X$ = $M$, $\Omega$, $\Sigma$, $\Delta$,
$\Phi$, of the operator rank $s$ = 0, 1, $\dots$, $2j_\chi$ and of the
parameter $l$, related to the explicit power of the transferred
momentum $q$ appearing in the operator (possible values of $l$ are: $l$
= $s$ for $X=M$, $\Omega$; $l$ = $s$-1, $s$, $s$+1 for $X$= $\Sigma$,
$\Phi$; $l$ = $s$-1, $s$ for $X$ = $\Delta$). See~\cite{all_spins} for
the explicit definitions of the $\mathcal{O}_{X,s,l}$ operators.  For
$j_\chi\le$1 the relation between the set of Ref.~\cite{all_spins} and
that of Refs.~\cite{haxton1,haxton2} (including there extension to
some spin-1 operators~\cite{krauss_spin_1,catena_krauss_spin_1}) is
listed in Table~\ref{tab:Haxton_operators}.

{ 
\begin{table}[t]\centering
  \caption{Non-relativistic Galilean invariant operators $\calO_j$
    introduced in~\cite{haxton2, krauss_spin_1, catena_krauss_spin_1})
    for a dark matter particle of spin $0$, $1/2$ and $1$, and their
    relation with the WIMP--nucleon operators $\calO_{X,s,l}$ defined
    in Ref.~\cite{all_spins}. }
\label{tab:Haxton_operators}
\renewcommand{\arraystretch}{1.5}
\addtolength{\tabcolsep}{2.0pt}
\vskip\baselineskip
\hspace{-4em}
\begin{minipage}{0.4\textwidth}
\begin{tabular}{@{}llr@{}}
\toprule
$\calO_{1}$ & $1$ & $\calO_{M,0,0}$ \\
[0.5ex]\cdashline{1-3}
$\calO_{2}$ & $(\vec{v}{}^{\plus}_{\chi N})^2 $ & $N.A.$ \\
[0.5ex]\cdashline{1-3}
$\calO_{3}$ & $-i \vec{S}_N \cdot ( \vec{\q} \times \vec{v}{}^{\plus}_{\chi N} ) $ & $-\calO_{\Phi,0,1}$ \\
[0.5ex]\cdashline{1-3}
$\calO_{4}$ & $\vec{S}_\chi \cdot \vec{S}_N$ & $\calO_{\Sigma,1,0}$ \\
[0.5ex]\cdashline{1-3}
$\calO_{5}$ & $ - i \vec{S}_\chi \cdot ( \vec{\q} \times \vec{v}{}^{\plus}_{\chi N} )$ & $-\calO_{\Delta,1,1}$ \\
[0.5ex]\cdashline{1-3}
$\calO_{6}$ & $(\vec{S}_\chi\cdot \vec{\q})  ( \vec{S}_N \cdot \vec{\q} )$ & $-\calO_{\Sigma,1,2}$ \\
[0.5ex]\cdashline{1-3}
$\calO_{7}$ & $ \vec{S}_N \cdot \vec{v}{}^{\plus}_{\chi N} $ & $\calO_{\Omega,0,0}$ \\
[0.5ex]\cdashline{1-3}
$\calO_{8}$ & $\vec{S}_\chi \cdot \vec{v}{}^{\plus}_{\chi N}$ & $\calO_{\Delta,1,0}$ \\
[0.5ex]\cdashline{1-3}
$\calO_{9}$ & $-i \vec{S}_\chi \cdot (\vec{S}_N \times \vec{\q} )$ & $\calO_{\Sigma,1,1}$ \\
[0.5ex]\cdashline{1-3}
$\calO_{10}$ & $-i \vec{S}_N \cdot \vec{\q}$ & $-\calO_{\Sigma,0,1}$ \\
[0.5ex]\cdashline{1-3}
$\calO_{11}$ & $-i \vec{S}_\chi \cdot \vec{\q}$ & $-\calO_{M,1,1}$ \\
[0.5ex]\cdashline{1-3}
$\calO_{12}$ & $\vec{S}_\chi \cdot (\vec{S}_N \times \vec{v}{}^{\plus}_{\chi N} )$ & $-\calO_{\Phi,1,0}$ \\
\bottomrule
\end{tabular}
\end{minipage}
\hspace{1em}
\begin{minipage}{0.4\textwidth}
\begin{tabular}{@{}llr@{}}
\toprule
$\calO_{13}$ & $\calO_{10}\calO_{8}$ & $-\calO_{\Phi,1,1}$ \\
[0.5ex]\cdashline{1-3}
$\calO_{14}$ & $\calO_{11}\calO_{7}$ & $-\calO_{\Omega,1,1}$ \\
[0.5ex]\cdashline{1-3}
$\calO_{15}$ & $-\calO_{11}\calO_{3}$ & $-\calO_{\Phi,1,2}$ \\
[0.5ex]\cdashline{1-3}
$\calO_{16}$ & $-\calO_{10}\calO_{5}$ &$-\calO_{\Phi,1,2}- \tilde{q}^2 \calO_{\Phi,1,0}$\\
[0.5ex]\cdashline{1-3}
$\calO_{17}$ & $-i\vec{\tilde{q}}\cdot{\bf \cal S}\cdot\vec{v}{}^{\plus}_{\chi N}$ & $\calO_{\Delta,2,1}$\\
[0.5ex]\cdashline{1-3}
$\calO_{18}$ & $-i\vec{\tilde{q}}\cdot{\bf \cal S}\cdot\vec{S}_N$ & $\calO_{\Sigma,2,1}-\frac{1}{3}\calO_{\Sigma,0,1}$\\
[0.5ex]\cdashline{1-3}
$\calO_{19}$ & $\vec{\tilde{q}}\cdot{\bf \cal S}\cdot\vec{\tilde{q}}$ & $\calO_{M,2,2}+\frac{1}{3}\tilde{q}^2\calO_{M,0,0}$\\
[0.5ex]\cdashline{1-3}
$\calO_{20}$ & $\left (\vec{S}_N \times \vec{\tilde{q}}\right )\cdot{\bf \cal S}\cdot\vec{\tilde{q}}$ & -$\calO_{\Sigma,2,2}$\\
[0.5ex]\cdashline{1-3}
$\calO_{21}$ & $\vec{v}{}^{\plus}_{\chi N}\cdot{\bf \cal S}\cdot\vec{S}_N$ & $\frac{1}{3}{\cal O}_{\Omega,0,0}$\\
[0.5ex]\cdashline{1-3}
$\calO_{22}$ & $\left (- i\vec{\tilde{q}}\times\vec{v}{}^{\plus}_{\chi N}\right )\cdot{\cal S}\cdot \vec{S}_N$ & $- {\cal O}_{\Phi,2,1}-\frac{1}{3}{\cal O}_{\Phi,0,1}$\\
[0.5ex]\cdashline{1-3}
$\calO_{23}$ & $- i\vec{\tilde{q}}\cdot{\cal S}\cdot\left (\vec{S}_N\times\vec{v}{}^{\plus}_{\chi N} \right )$ & $- {\cal O}_{\Phi,2,1}+\frac{1}{3}{\cal O}_{\Phi,0,1}$\\
[0.5ex]\cdashline{1-3}
$\calO_{24}$ & $- \vec{v}{}^{\plus}_{\chi N}\cdot{\cal S}\cdot\left (\vec{S}_N\times i\vec{\tilde{q}}\right)$ & $- {\cal O}_{\Phi,2,1}-\frac{1}{3}{\cal O}_{\Phi,0,1}$\\
\bottomrule
\end{tabular}
\vspace{1.2\baselineskip}
\end{minipage}
\end{table}
} 

As a consequence, besides an integer, each key $n$ of the dictionary {\verb wc }
can also be a tuple {\verb (X,s,l) } with {\verb X } a
string among {\verb 'M' }, {\verb 'Omega' }, {\verb 'Sigma' }, {\verb 'Delta' }
and {\verb 'Phi' } and {\verb s }, {\verb l } two integers
following the restrictions explained above\footnote{WimPyDD
  does not allow to mix the two different sets of operators in the same
  Hamiltonian.}.

\section{Initializing the experiment directory}
\label{app:exp}

The calculation of the integrated functions of
Eq.~(\ref{eq:rbar}), which include the effect of the detector's response, 
 require to fix the target $T$, the energy intervals
$[E^{\prime}_1,E^{\prime}_2]$, as well as the differential response
functions $r_{T}^i(E_R, E^{\prime})$ of Eq.~(\ref{eq:differential_r}).
The latter depend on the exposure $MT_0$, the efficiency
$\epsilon(E^{\prime})$, the energy resolution ${\cal
  G}(E^{\prime},E_{ee})$ and the quenching factors $q_T(E_R)$ (for each
element in the target).  As explained in Section~\ref{sec:load_response_function}
at instantiation the {\verb experiment } class
stores each of these elements in an appropriate attribute, reading the
information relevant to each of them from a separate file contained in
a sub--folder of the {\verb WimPyDD/Experiments }
folder of the working directory. The name of the subfolder is passed as an
argument to the {\verb experiment } class. The list of files is given in
Table~\ref{table:files}. For all of them if the file is missing the default
value indicated in the Table is adopted. In particular the target material
has no default and can be either passed as a {\verb target }
object (see~\ref{sec:element}) to the {\verb experiment } class, or
provided in the file {\verb target.tab } as a valid input string
for the {\verb target } class. 

All input files must contain plain text. In the following we review
each of them.  All the lines of a {\verb .tab } file
that do not match the expected format are ignored with the exception
of the first, that is interpreted as a help string. The {\verb .py }
files are loaded as ordinary libraries, and must contain a valid
python function implementing the relevant quantity (energy resolution,
efficiency, quenching). Such function can have an arbitrary name (WimPyDD loads
the first function defined in the file) and its user--provided help
(within {\verb '''...''' } triple quotation marks) is loaded as a help string.
All help strings are printed issuing a {\verb print } command of the
{\verb experiment } object. A few examples of experiment directories are provided
with the WimPyDD distribution. 

\begin{table}[t]\centering
  \caption{Files in the experiment subfolder containing the information
    to calculate the response functions. \label{table:files}}
\renewcommand{\arraystretch}{1.1}
\addtolength{\tabcolsep}{2.0pt}
\vskip\baselineskip
\begin{tabular}{@{}ccc@{}}
\toprule
{\bf Physical quantity(unit)} & {\bf Default(if missing file)}& {\bf File name}  \\
\midrule
Target material & no default & {\verb target.tab }\\
\midrule
$MT_0$(kg day) & 1 & {\verb exposure.tab } \\
\midrule
$\epsilon(E^{\prime})$ & 1 & {\verb efficiency.tab } or \\
$      $ &   & {\verb efficiency.py } \\
\midrule
${\cal G}(E^{\prime},E_{ee})$ & $\delta(E^{\prime}-E_{ee})$ & {\verb resolution.py }\\
\midrule
$q_T(E_R)$ & 1 & {\verb <element_name>_quenching.py } or\\
$      $ &   &  {\verb <element_name>_quenching.tab } \\
$      $ &   &  (ex: {\verb xenon_quenching.tab }) \\
\midrule
$[E^{\prime}_1,E^{\prime}_2]$(keV) & $E^{\prime}_1$=$E^{\prime}_2$=1 keV &  {\verb data.tab }\\
$      $ & (differential rate)  &   \\
\midrule
Modifier  & 1 & {\verb modifier.tab }\\
(see Appendix~\ref{app:modifier_file}) & &\\
\bottomrule
\end{tabular}
\label{tab:exp_file_list}
\end{table}

\subsection{Target file}
\label{app:target_file}
If present the target file must contain a string with a valid input for the
{\verb WD.target } class, for instance {\verb Ge }, {\verb CF3I }
or {\verb NaI } (no quotation marks). If the {\verb target.tab } file is not
provided a {\verb target } object should be passed as a parameter
to the {\verb experiment } class.
In this case the corresponding {\verb target.tab } file will be added to
the experiment directory. WimPyDD will prompt the user for an input
if neither a {\verb target } object is passed to the {\verb experiment }
class nor a {\verb target.tab } is provided. Moreover it will produce
a warning if the user passes a {\verb target } object to the
{\verb experiment } class that does not match the content of an
existing {\verb target.tab } file in the experiment directory.

\subsection{Exposure file}
\label{app:exposure}

This file can contain either a float with the exposure $M T_0$ in kg/day or
a list/array with a different exposure for each of the experimental bins
contained in {\verb data.tab }, and controls the output of the
{\verb wimp_dd_rate } routine. Setting the content of
{\verb exposure.tab } to 1 (or, equivalently, if the {\verb exposure.tab }
file is missing) will produce an output normalized to events/kg/day.
Moreover an output in events/kg/day/keV will be obtained by providing
in {\verb exposure.tab } a list or array with the
inverses of the energy bin widths contained in {\verb data.tab }.

\subsection{Efficiency and quenching files}
\label{app:efficiency_quenching}
These files must contain either two columns with a sampling of the arguments and 
values of the function to be used for interpolation
(a simple linear interpolation is used) or a valid Python function.
Notice that the efficiency function $\epsilon$ must depend on the
visible energy $E^{\prime}$ (in keV) while the quenching
(or light yield) $Q(E_R)$ must depend on the recoil energy $E_R$
(also in keV). A different file for the quenching factor of each element in the target
can be provided, with the naming convention of Table~\ref{tab:exp_file_list}.

\subsection{Energy resolution file}
\label{app:energy_resolution}
Since the function ${\cal G}(E^{\prime},E_{ee})$ takes two arguments only a
{\verb .py } file can be provided with a regular Python function in the two
variable $E^{\prime}$ (visible energy) and $E_{ee}$ (electron-equivalent energy)
in keV:

\begin{Verbatim}[frame=single,xleftmargin=1cm,xrightmargin=1cm,commandchars=\\\{\}]
def resolution(e_prime,e_ee):
    ...
\end{Verbatim}

\subsection{Data file}
\label{app:data}
The file {\verb data.tab } must contain the energy bins 
$[E^{\prime}_1,E^{\prime}_2]$ needed for the calculation
of the response function of~Eq.~(\ref{eq:rbar})
that affects the output of the {\verb wimp_dd_rate } routine.
Each line of the {\verb data.tab } file corresponds
to a different energy bin and can be in different format, as summarized
in Table~\ref{table:data_formats}. In particular, if a line contain a single float
{\verb wimp_dd_rate } will return the differential
rate at the corresponding value of the energy and the response functions of
Eq.~(\ref{eq:rbar}) will not be integrated upon $E^{\prime}$
(the differential rate can be calculated more efficiently using the
{\verb diff_rate } routine, although without the effect of the detector's response). 
On the other hand two floats will be interpreted as
an interval of visible energy $E^{\prime}$.
If present, the information following the two floats of
an energy bin  is used by {\verb mchi_vs_exclusion } to
calculate a 90\% C.L. upper
bound: in this case a single integer will be interpreted as a
measured number of counts
with Poissonian fluctuation, two floats as a number of counts with
the corresponding
Gaussian error. Each line of the {\verb data.tab } file
can have different format (single energy value/energy interval,
with/without upper bound information). 
Bins without upper bound information are neglected by the
{\verb mchi_vs_exclusion } routine.

\subsection{Modifier}
\label{app:modifier_file}
If present, the file {\verb modifier.py } must contain a function

\begin{Verbatim}[frame=single,xleftmargin=1cm,xrightmargin=1cm,commandchars=\\\{\}]
f(er,eprime,exp,n_element,n_isotope,n_bin)
\end{Verbatim}

\noindent with {\verb er }=$E_R$, {\verb eprime }=$E^{\prime}$, {\verb exp } an object
belonging to the {\verb experiment } class, {\verb n_element } and {\verb n_isotope }
two integers that specify the element in the {\verb exp.target.element } array and
the isotope, respectively and {\verb n_bin } an integer identifying
the energy bin (among those defined in {\verb data.tab }). The default for the function
is unity. At the time of calculating the integrated response functions
$\left [ \bar{{\cal R}}^a_{T,[E_1^{\prime},E_2^{\prime}]}(E_R) \right ]_{jk}^{\tau\tau^{\prime}}$
the differential response functions\\
$\left [r^a_T(E_R,E^\prime)\right ]_{jk}^{\tau\tau^{\prime}}$ are multiplied times the modifier.
Its purpose is to implement
specific cases where the expected rate is not in the form of
Eqs.~(\ref{eq:start1},\ref{eq:dr_der}), allowing, in principle, a different
modifier for each target or energy bin. For instance, droplet detectors and bubble
chambers such as COUPP~\cite{coupp}, PICO60~\cite{pico60_2019} or PICASSO~\cite{PICASSO} 
are threshold detectors
(i.e. a signal is detected only above some value of the deposited
energy, which is fixed by changing the running parameters such as the
temperature or the pressure). In this case the expected number of
events is given by:

\begin{equation}
R=N_T MT\int_0^{\infty} P(E_R) \frac{dR}{dE_R} dE_R,
\label{eq:r_threshold}
\end{equation}

\noindent where $P(E_R)$ is the probability that the energy $E_R$ will
produce a signal if the energy threshold is fixed to $E_{th}$. The
response function for such class of detectors requires that no energy
resolution is provided while the $P(E_R)$ function
is implemented through the file {\verb modifier.py }.

\begin{table}[t]\centering
  \caption{Accepted format of each line of the \protect {\tt data.tab }file.
    \label{table:data_formats}}
\renewcommand{\arraystretch}{1.5}
\addtolength{\tabcolsep}{2.0pt}
\vskip\baselineskip
\begin{tabular}{@{}cc@{}}
\toprule
Line content & {\verb WimPyDD } interpretation \\ 
\midrule
    {\verb 10 } & $E^{\prime}$=10 keV  \\
    {\verb 10 }\, {\verb 20 } & $[E_1^{\prime},E_2^{\prime}]$=$[$10,20$]$ keV  \\
    {\verb 10 }\, {\verb 20 }\, {\verb 1 } & $[E_1^{\prime},E_2^{\prime}]$=$[$10,20$]$ keV, $N_{counts}$=1 events  \\
    {\verb 10 }\, {\verb 20 }\, {\verb 3.2 } \, {\verb 0.4 } & $[E_1^{\prime},E_2^{\prime}]$=$[$10,20$]$ keV, $N_{counts}$=3.2$\pm$0.4  events  \\
\midrule
\bottomrule
\end{tabular}
\end{table}


\section{Adding new nuclear targets}
\label{app:add_elements}

Element targets are instantiated by the {\verb element } class, that
takes as an argument a string with the element symbol. By convention
element symbols must begin with a capital letter. The string must
correspond to a file in the directory {\verb WimPyDD/Targets }
containing the required information. For instance the xenon element,
with symbol {\verb Xe }, is instantiated with:

\begin{Verbatim}[frame=single,xleftmargin=1cm,xrightmargin=1cm,commandchars=\\\{\}]
  xenon=WD.element('Xe')
\end{Verbatim}

\noindent and the class requires the text file {\verb "Xe.tab" } in
{\verb WimPyDD/Targets }. The content of such file is the following:

\begin{Verbatim}[frame=single,xleftmargin=1cm,xrightmargin=1cm,commandchars=\\\{\}]
xenon
xe
54
128     0.01910 0       11\\
129     0.26401 0.5     12\\
130     0.04071 0       13\\
131     0.21232 1.5     14\\
132     0.26909 0       15\\
134     0.10436 0       16\\
136     0.08857 0       17
\end{Verbatim}

\noindent The first three
lines contain the element full name, symbol, and atomic
number $Z$. For each of the element isotopes each of the following
lines initializes the atomic mass number $A$ in the first column,
the fractional abundance in the second column and the spin $j_T$ in
the third column. Finally the last column contains the internal
integer code {\verb itar }=1,$\dots$, 30 which is used to initialize the
nuclear response functions $W_{l,T}(q)$ provided in
Refs.~\cite{haxton2, catena} for the 30 isotopes
$T$=$^{12}C$, $^{19}F$, $^{23}Na$, $^{28}Si$, $^{70}Ge$,$^{72}Ge$,
$^{73}Ge$, $^{74}Ge$, $^{76}Ge$, $^{127}I$, $^{128}Xe$, $^{129}Xe$,
$^{130}Xe$, $^{131}Xe$, $^{132}Xe$, $^{134}Xe$, $^{136}Xe$, $^{16}O$,
$^{40}Ar$, $^{40}Ca$, $H$, $^{3}He$, $^{4}He$, $^{14}N$, $^{20}Ne$,
$^{24}Mg$, $^{27}Al$, $^{32}S$, $^{56}Fe$, $^{59}Ni$. In particular,
the following parameterization is used~\cite{haxton2}:

\begin{equation}
 W_{l,T}^{\tau,\tau^\prime}(q) e^{-2y}\sum_n a^{\tau,\tau^\prime}_n y^n,
  \label{eq:w_coeff}
\end{equation}

\noindent where $y$ = $(qb/2)^2$, $b$ = $\sqrt{41.467/(45
  A^{-1/3}-25A^{-2/3})}$. The $a_n^{\tau\tau^\prime}$ coefficients
are contained at the line $8(\mbox{itar}-1)+l+\tau+\tau^\prime$ of the file

\begin{Verbatim}[frame=single,xleftmargin=1cm,xrightmargin=1cm,commandchars=\\\{\}]
  WimPyDD/Targets/Nuclear_response_functions/\\
  nuclear_response_functions_coefficients_table.dat 
\end{Verbatim}

\noindent where $l=1,...8$ corresponds to the 8 nuclear response functions $M$,
$\Sigma^{\prime\prime}$, $\Sigma^{\prime}$, $\Phi^{\prime\prime}$,
$\tilde{\Phi}^{\prime}$, $\Delta$, $\Phi^{\prime\prime} M$,
$\Delta \Sigma^{\prime}$). Upon instantiation the element
object acquires the method {\verb func_w }, an array that contains
a function  of the single argument {\verb q } for each isotope of the element.
Such function returns
an array  with shape (8,2,2)=($l$,$\tau$,$\tau^\prime$) with the values of
the nuclear response functions of Eq.~(\ref{eq:w_coeff}).

For new elements or isotopes for which numerical calculations of the
$W_{l,T}^{\tau,\tau^\prime}(q)$ functions are not available the internal
code {\verb itar } must be set to zero. 
It is possible to provide the corresponding nuclear response function (or
to override the
default definitions of~\cite{haxton2, catena}) adding the file
{\verb <isotope_name>_func_w.py } in the directory  \newline 
{\verb WimPyDD/Targets/Nuclear_response_functions } with {\verb <isotope_name> } 
the isotope name, that must match one of the entries of the 
 {\verb isotopes } attribute of the {\verb element } class. Such file must contain a custom 
 definition of the nuclear response function
with the same behaviour of the default one, i.e. it must
depend on the single argument {\verb q } and return an array
with shape (8,2,2). For instance, in order to change $W_{l,T}$ for
$T$=$^{129}Xe$ the custom definition must be contained in the file
{\verb 129Xe_func_w.py } with {\verb '129Xe' } one of the entries of the {\verb xenon.isotopes }
 array (a few examples of custom definition files
are provided with the WimPyDD distribution).  If {\verb itar=0 } and no
file is present the corresponding  nuclear response
functions are set to zero.

\section{The halo function}
\label{sec:halo}
WimPyDD contains a routine {\verb streamed_halo_function } that calculates the
$v_k$, $\delta\eta_k^{(0,1)}$ quantities needed by the {\verb diff_rate } and
{\verb wimp_dd_rate }. The simple call:

\begin{Verbatim}[frame=single,xleftmargin=1cm,xrightmargin=1cm,commandchars=\\\{\}]
  vmin,delta_eta0=WD.streamed_halo_function()
\end{Verbatim}

\noindent samples linearly the $\delta\eta_k^{(0)}$'s for a Maxwellian
velocity distribution cut at the escape velocity in the Earth's rest
frame:

\begin{eqnarray}
  f(\vec{v},t)&=& N\left(\frac{3}{ 2\pi v_{rms}^2}\right )^{3/2}
  e^{-\frac{3|\vec{v}+\vec{v}_{E,Gal}|^2}{2 v_{rms}^2}}\Theta(v_{esc,Gal}-|\vec{v}+\vec{v}_{E,Gal}(t)|),\\
  N &=& \left[erf(z)-\frac{2}{\sqrt{\pi}}z e^{-z^2}\right]^{-1},  
  \label{eq:maxwellian}
  \end{eqnarray}

\noindent with $v_{esc,Gal}$ the escape velocity in the Galactic rest
frame and $z^2$ = $3 v_{esc,Gal}^2/(2 v_{rms}^2)$. By default hydrothermal
equilibrium between the WIMP gas pressure and gravity is assumed,
$v_{rms}$=$\sqrt{3/2}|\vec{v}_{rot,Gal}|$ with
$\vec{v}_{rot,Gal}$ the galactic rotational velocity (such value can be modified
using the argument {\verb vrms }). Moreover
$\vec{v}_{E,Gal}$ = $\vec{v}_{rot,Gal}$+ $\vec{v}_{Sun,rot}$ +
$\vec{v}_{E,Sun}$, with $\vec{v}_{Sun,rot}$ the peculiar velocity of
the solar system with respect to the Galactic rotational curve and
$\vec{v}_{Sun,rot}$ the velocity of the Earth around the Sun.  The
routine assumes the default values $v_{esc,Gal}$ = 550 km/s,
$\vec{v}_{rot,Gal}$ = $(0,220,0)$ km/s, $\vec{v}_{Sun,rot}$ = $(9,12,7)$
km/s (in the Galactic reference frame where the $z$ axis is
in the direction perpendicular to the galactic plane, the $x$ axis
points toward the galactic center and the velocity of the solar system
points in the $y$ direction). Such values can be changed by using the arguments
{\verb v_esc_gal }, {\verb v_rot_gal } and {\verb v_sun_rot }, respectively.
For $\vec{v}_{E,Sun}$ the expression in~\cite{Lewin_Smith_1996} is used.
To calculate $\delta\eta^{(1)}$ the argument {\verb yearly_modulation=True } must be used,
and to get $\eta^{(0,1)}_k$ instead of $\delta\eta^{(0,1)}_k$ use {\verb delta_eta=False }.

The routine can also be used for a user--defined velocity
distribution.  This requires to pass the argument {\verb velocity_distribution_gal }.
For instance, setting:

\begin{Verbatim}[frame=single,xleftmargin=1cm,xrightmargin=1cm,commandchars=\\\{\}]
  
def func(u,sigma_x,sigma_y,sigma_z):\\
      return np.exp(-0.5*(u[0]**2/sigma_x**2+\\
      u[1]**2/sigma_y**2+u[2]**2/sigma_z**2))\\
\\
\end{Verbatim}

\noindent the corresponding time-averaged $\delta \eta_k^{(0)}$'s can be calculated by:

\begin{Verbatim}[frame=single,xleftmargin=1cm,xrightmargin=1cm,commandchars=\\\{\}]
vmin,delta_eta_0=WD.streamed_halo_function(\\
velocity_distribution_gal=func,\\
sigma_x=120, sigma_y=150, sigma_z=170)
\end{Verbatim}

\noindent The user--define velocity distribution {\verb func } 
must be a function of a three--dimensional array containing the WIMP
velocity in the galactic rest frame. The parameters of {\verb func }
are passed with the keyworded variable-length argument list
{\verb **args } (the argument {\verb vrms } for the Maxwellian distribution is just a
  particular case of this). All the other parameters of the routine work in the same way
  as for a Maxwellian.

  Both for a Maxwellian and for a user--defined velocity distribution
  the full expressions of Eqs.~(\ref{eq:eta0}) and (\ref{eq:eta1})
  (discretized as sums over the days of the year) can be used
  by setting {\verb full_year_sample=True }. This can be time consuming, so that by default
  {\verb full_year_sample } = {\verb False }. In this case $\eta^{(0)}$ is
  approximated by setting $\vec{v}_{E,Sun}\rightarrow$0. As far as the modulated 
  component  $\eta^{(1)}$  is concerned, in the Maxwellian case it is approximated by using a linear expansion in the parameter
  $|\vec{v}_{E,Sun}|/|\vec{v}_{E,Gal}|$, while for the case of a user--defined velocity distribution it is calculated using the expression $[\eta(t=t_0)-\eta(t=t_0+T/2)]/2$. Also $t_0$ can be changed by using the
  argument {\verb modulation_phase } in days. Finally, {\verb day_of_the_year } allows
  to calculate the halo function on a specific day of the year.

  In the case when {\verb velocity_distribution_gal } is used
  and the routine calculates numerical integrals the output is
  saved in the folder {\verb Halo_functions } for later use to speed-up
  successive evaluations of the same halo function, unless
  {\verb recalculate=True }.
\section{Summary of WimPyDD components}
\label{app:summary}

In table~\ref{table:wimpydd_components} we provide a short summary of
the main components of the WimPyDD code with their correspondence
to the Equations in Section~\ref{sec:theory} and Appendix~\ref{app:response_functions}.
To learn more about each of them use the Python help function.
\begin{table}[t]\centering
  \caption{List of the main components of the WimPyDD code. \label{table:wimpydd_components}}
\addtolength{\tabcolsep}{1.0pt}
\vskip\baselineskip
\begin{tabular}{@{}cccc@{}}
\toprule
{\bf Section} &  & {\bf Equation} & {\bf WimPyDD}  \\
\midrule
\ref{app:response_functions}&$\left( \frac{d\sigma}{d E_R}\right )_T$&(\ref{eq:dsigma_der}) & {\verb dsigma_der }\\
\midrule
\ref{app:response_functions}&$\frac{1}{2j_{\chi}+1} \frac{1}{2j_{T}+1}\sum_{spin}|{\cal M}(q)|^2_T$&(\ref{eq:haxton_40}) & {\verb eft_amplitude_squared }\\ 
\midrule
\ref{app:response_functions}&$W_{l,T}^{\tau\tau^{\prime}}$&(\ref{eq:haxton_40}) & {\verb target.element[i].func_w }\\ 
\midrule
\ref{app:response_functions}&$\delta\eta_k$&(\ref{eq:delta_eta_piecewise}) & {\verb streamed_halo_funtion }\\ 
\midrule
\ref{app:response_functions}&$\left(\frac{dR}{dE_R}\right)_T$&(\ref{eq:dr_der_piecewise}) & {\verb diff_rate }\\
\midrule
\ref{app:response_functions}&$S^{0,1}_{[E_1^{\prime},E_2^{\prime}]}$&(\ref{eq:s0}) and~(\ref{eq:sm}) & {\verb wimp_dd_rate }\\ 
\midrule
\ref{sec:theory}&$\mathcal{H}=\sum_{\tau=0,1} \sum_{j} c_j^{\tau}\mathcal{O}_{j}t^{\tau}$&(\ref{eq:H}) & {\verb eft_hamiltonian }\\
\midrule
\ref{sec:theory}&$\left [ \bar{{\cal R}}^a_{T,[E_1^{\prime},E_2^{\prime}]}(E_R) \right ]_{jk}^{\tau\tau^{\prime}}$&(\ref{eq:r_factorization}) & {\verb experiment.response_functions }\\ 
\bottomrule
\end{tabular}
\end{table}



\end{document}